\newif\ifPDF
\ifx\pdfoutput\undefined\PDFfalse
\else\ifnum\pdfoutput > 0\PDFtrue
     \else\PDFfalse
     \fi
\fi
\documentclass[useAMS,usenatbib]{mn2e}
\ifPDF
  \usepackage[T1]{fontenc}
  \usepackage{aeguill}
  \usepackage[pdftex]{graphicx,color}
  \usepackage{subfigure}
  \usepackage{afterpage}
\else
  \usepackage[T1]{fontenc}
  \usepackage[dvips]{graphicx,color}
  \usepackage{subfigure}
  \usepackage{afterpage}
\fi
\usepackage{times}

\usepackage{graphics}
\usepackage{psfig}
\usepackage{rotating}
\usepackage{multirow}

\def\sb{{\sb}}

\title[Near-IR studies of WR stars in 
Westerlund~1] {A census of the Wolf-Rayet content in Westerlund~1 from
near-infrared imaging and spectroscopy\thanks{Based on observations made
with ESO telescopes at the La Silla Observatory under programme IDs 
073.D-0321 and 075.D-0469}}

\author[Crowther et al.]{Paul A. Crowther$^{1}$, L. J. Hadfield$^{1}$, 
J. S. Clark$^{2}$, I. Negueruela$^{3}$ and W. D. Vacca$^{4}$\\
$^{1}$ Department of Physics \& Astronomy, 
University of Sheffield, Hicks Building, Hounsfield Rd, 
Sheffield, S3 7RH, UK\\
$^{2}$ Department of Physics \&  Astronomy, The Open University,
Milton Keynes, MK7 6AA, UK\\
$^{3}$ Dpto de Fisica, Ingenieria de Sistemas y Teoria de la Senal, 
Universidad de Alicante, Apdo. 99, E03080, Alicante, Spain\\
$^{4}$ SOFIA-URSA, NASA Ames Research Center, MS N211-3, Moffett Field, 
CA~94035, USA
}

\begin{document}

\date{Accepted 2006 Aug 16; Received 2006 August 2; in original form 
2006 Jul 21}

\pagerange{\pageref{firstpage}--\pageref{lastpage}} \pubyear{2006}

\maketitle

\label{firstpage}

\begin{abstract} New NTT/SOFI imaging and spectroscopy of the
Wolf-Rayet population in Westerlund~1 are presented. Narrow-band near-IR
imaging
together with follow up spectroscopy reveals four new Wolf-Rayet stars, of 
which three were independently identified recently by Groh et al., 
bringing the confirmed  Wolf-Rayet content to 24 (23 excluding source S) 
-- representing 8\% of the known Galactic Wolf-Rayet population --
comprising 8 WC stars
and 16 (15) WN stars. Revised coordinates and near-IR photometry are
presented, whilst a quantitative near-IR spectral classification scheme 
for Wolf-Rayet stars is presented and applied to members of Westerlund~1. 
Late subtypes are dominant, with no subtypes earlier than 
WN5 or WC8 for the nitrogen and carbon sequences, respectively. 
A qualitative inspection of the WN
stars suggests that most ($\sim$75\%) are highly H-deficient. The Wolf-Rayet
binary fraction is high ($\geq$62\%), on the basis of dust emission from 
WC stars, 
in addition to a significant WN binary fraction from hard X-ray detections 
according to Clark et al. We exploit the large WN population of Westerlund~1 to reassess its
distance ($\sim$5.0~kpc) and extinction ($A_{K_{S}} \sim$ 0.96 mag),
such that it is located at the edge of the Galactic bar, with an oxygen metallicity 
$\sim$60\% higher than Orion. The observed ratio of WR stars to red and 
yellow hypergiants, 
N(WR)/N(RSG+YHG) $\sim$ 3, favours an age of $\sim$4.5--5.0 Myr, with individual 
Wolf-Rayet stars descended from progenitors of initial mass $\sim 40-55 M_{\odot}$.
Qualitative estimates of current masses for non-dusty, H-free WR stars are presented,
revealing $10-18 M_{\odot}$, such that $\sim$75\% of the initial
stellar mass has been removed via stellar winds or close binary evolution.
We present a revision to the cluster turn-off mass for other Milky 
Way  clusters in which Wolf-Rayet stars are known, based upon the 
latest temperature calibration for OB stars. Finally, comparisons
between the observed WR population and subtype distribution 
in Westerlund~1 and instantaneous burst evolutionary synthesis models are 
presented.
\end{abstract}


\section{Introduction}

Several hundred Wolf-Rayet (WR) stars -- the evolved descendants of the
most massive O stars -- have been identified within the Milky Way 
(van der 
Hucht 2006). In principle,
studies of WR stars in open clusters 
provides excellent
observational constraints upon their ages and initial masses (e.g.
Schild \& Maeder 1984). In practice, this has proved challenging, due
to the small number of WR stars observed within individual Milky Way or 
Magellanic Cloud clusters. This is unsurprising, given their short 
lifetimes and small number of suitably massive stars in typical open
clusters ($\approx 10^{3} M_{\odot}$), for which empirical
studies indicate a maximum stellar mass of $\sim 30-35 M_{\odot}$ 
(e.g. Weidner \&  Kroupa 2006).
Hitherto, solely the 
Arches, Quintuplet and Galactic Centre clusters have offered the potential 
for studying large numbers of WR stars in suitably massive ($10^{4} 
M_{\odot}$) clusters, for which significantly higher mass stars
are observed, albeit hindered by exceptionally high interstellar 
extinction.

Fortunately, Westerlund~1 (Westerlund 1961) offers the 
possibility of undertaking a comprehensive study of WR stars at Solar 
or moderately super-Solar metallicity, since its mass has been estimated at 
10$^{5} M_{\odot}$ (Clark et al. 2005).  
Instantaneous burst evolutionary synthesis models at Solar metallicity 
(e.g. Starburst99, Leitherer et al. 1999) predict 
20--30 WR stars at an age of 4.5~Myr for such 
a high cluster mass. Observationally,  Clark \&  Negueruela (2002),  
Negueruela  \& Clark (2005) and Hopewell et al. (2005)  have identified 
20   WR stars within 4 arcmin ($\sim$6 pc at a distance of 5~kpc) of the 
central cluster, most of which were discovered  serendipitously. 

\begin{table*}
\caption{Catalogue of Wolf-Rayet stars in Westerlund~1,
including WR nomenclature following van der Hucht (2006),
following previous studies by Clark \& Negueruela (2002, CN02),
Negueruela \& Clark (2005, NC05), Hopewell et al. (2005,
H05), Negueruela (2005, priv comm, N05) and Groh et al. (2006, G06).
Photometry is obtained primarily from NTT/SOFI or 2MASS 
(Skrutskie et al. 2006).
Coordinates are obtained from our NTT/SOFI images, except
where noted, based upon astrometry of A--G from 3.6cm
radio images (see CN02). Previously published coordinates for I and J 
were in error due to spatial crowding.}
\label{WRcat}
\begin{tabular}{
c@{\hspace{2mm}}
l@{\hspace{3mm}}
l@{\hspace{2mm}}
l@{\hspace{3mm}}
l@{\hspace{2mm}}
r@{\hspace{3mm}}
r@{\hspace{3mm}}
r@{\hspace{3mm}}
c@{\hspace{3mm}}
l@{\hspace{3mm}}
l@{\hspace{3mm}}
l}
\hline
Source & Alias & WR & $\alpha$ & $\delta$  & J & H & K$_{S}$ & 
Ref & Previous & Ref & This\\
               &       &    & \multicolumn{2}{c}{J2000} & mag & mag & 
mag & &  Sp Type  & & Study\\
\hline
A & Wd1-72 & WR77sc & 16 47 08.32 & --45 50 45.5 & 10.34 & 9.11: & 8.37: & 
SOFI & WN4--5, $<$WN7 & CN02, NC05 & WN7b \\  
B  &  & WR77o & 16  47 05.36 & --45 51 05.0 & 10.91 & 9.79 & 9.18 & SOFI
& WNL, WN8? & CN02, NC05 & WN7o \\
C &  & WR77m & 16 47 04.40 & --45 51 03.8 & 11.26 & 9.51 & 8.23 & SOFI & 
WC8, WC8.5 & CN02, NC05 & WC9d \\
D &  & WR77r & 16 47 06.24 & --45 51 26.5 & 11.63 & 10.31 & 9.61 & 
SOFI & 
WN6--8  & CN02 & WN7o \\
E & Wd1-241 & WR77p & 16 47 06.05 & --45 52 08.2 & 10.12 & 9.09 & 8.29: & 
SOFI & WC9  & CN02 & WC9 \\ 
F & Wd1-239 & WR77n & 16 47 05.22 & --45 52 25.0 & 9.85 &  7.97 & 7.28 & 
SOFI & WC9  & CN02 & WC9d \\
G &   & WR77j & 16 47 04.01 & --45 51 25.2 & 11.36 & 9.97 & 9.28 & 
SOFI &  WN6--8  & CN02 & WN7o \\
H &  & WR77l & 16 47 04.22 & --45 51 20.2 & 10.31 & 8.56 & 7.38 & 
SOFI & 
WC9  & CN02 & WC9d \\
I &     & WR77c & 16 47 00.88 & --45 51 20.8 & 10.89 & 9.57 & 8.86 & 
SOFI& WN6--8   & CN02 & WN8o \\
J &      & WR77e & 16 47 02.47 & --45 51 00.1 &11.7: & 10.3: & 9.7: & SOFI 
& WNL   & CN02 & WN5h \\ 
K &       & WR77g & 16 47 03.25 & --45 50 43.8 & 11.81 &  10.40 & 
 9.53 & SOFI & WC, WC7 & CN02, NC05 & WC8 \\
L & Wd1-44 & WR77k & 16 47 04.19 & --45 51 07.4 & 9.08 & 7.72 & 7.19 & 
SOFI & WN9   & NC05 & WN9h:\\
M & Wd1-66 & WR77i & 16 47 03.96 & --45 51 37.8 & 10.13 & 7.64 & 6.9: & 
SOFI& WC9  &  NC05 & WC9d \\ 
N &     & WR77b   & 16 46 59.9 & --45 55 26 &9.69 & 7.84 & 6.41 &2MASS &
WC8  & NC05 & WC9d\\
O &    & WR77sb & 16 47 07.66 & --45 52 35.9 & 11.00 & 9.98 & 9.45 & SOFI
& WN6   & NC05 & WN6o \\
P & Wd1-57c & WR77d & 16 47 01.59 & --45 51 45.5 & 11.06 & 9.83 & 9.26 & 
SOFI & WN8  & NC03 & WN7o \\
Q &      & WR77a  & 16 46 55.55 &--45 51 35.0&11.72& 10.67&10.00& 
SOFI & WN6--7  & NC05 & WN6o\\
R & Wd1-14c & WR77q & 16 47  06.07 & --45 50 22.6 & 11.92 & 10.84 & 10.26 
& SOFI & WN6--7   & NC05 & WN5o \\ 
S & Wd1-5 & WR77f & 16 47 02.98 & --45 50 20.0 & 9.81 & 8.80: & 8.29 & 
SOFI & WNVL  & NC05 & WN10-11h or B0--1Ia$^{+}$\\ 
T & HBD4 & WR77aa & 16 46 46.3 &--45 47 58 &10.04&8.21&6.72 &
2MASS & WC9d   & H05 & WC9d\\
U & \#1  & WR77s & 16 47 06.55 & --45 50 39.0 & 10.77 & 9.72 & 
9.20 & SOFI &
WN4, WN5--7  & N05, G06 & WN6o \\
V &  & WR77h & 16 47  03.81 & --45 50 38.8 &  10.75 & 9.42 & 
8.76 & SOFI & WN8  & N05 & WN8o \\
W & \#3 & WR77sa & 16 47 07.58 & --45 49 22.2 & 12.11 & 10.75 & 10.04 & 
SOFI
& WN5--6   & G06 & WN6h \\
X & \#2 & WR77sd & 16 47 14.1 & --45 48 32 & 12.36 & 11.08 & 10.25 & 2MASS  
& WN4--5  & G06 & WN5o \\
\hline
\end{tabular}
\end{table*}

\begin{figure}
\centerline{\psfig{figure=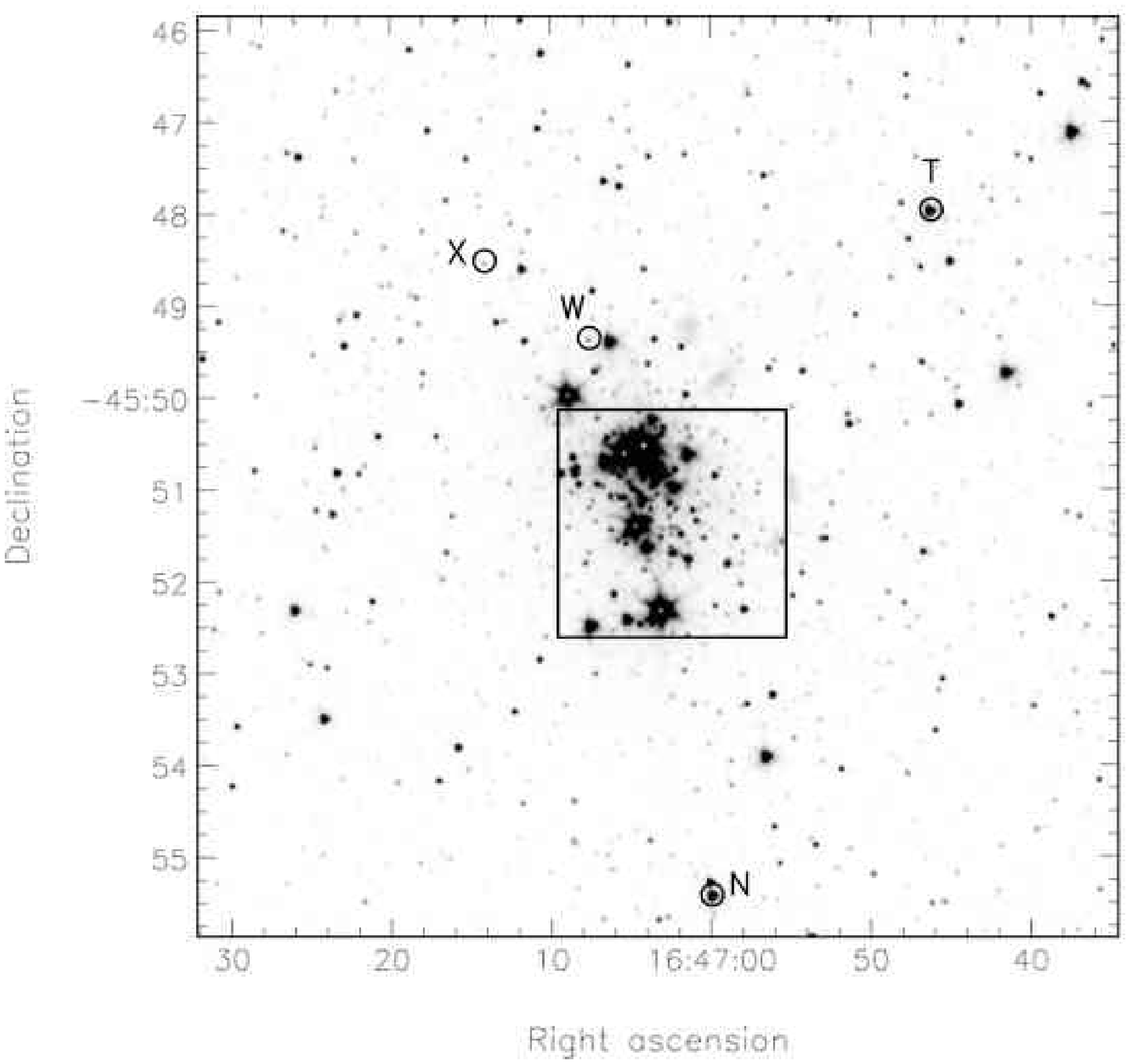,width=8.8cm,angle=0}}
\centerline{\psfig{figure=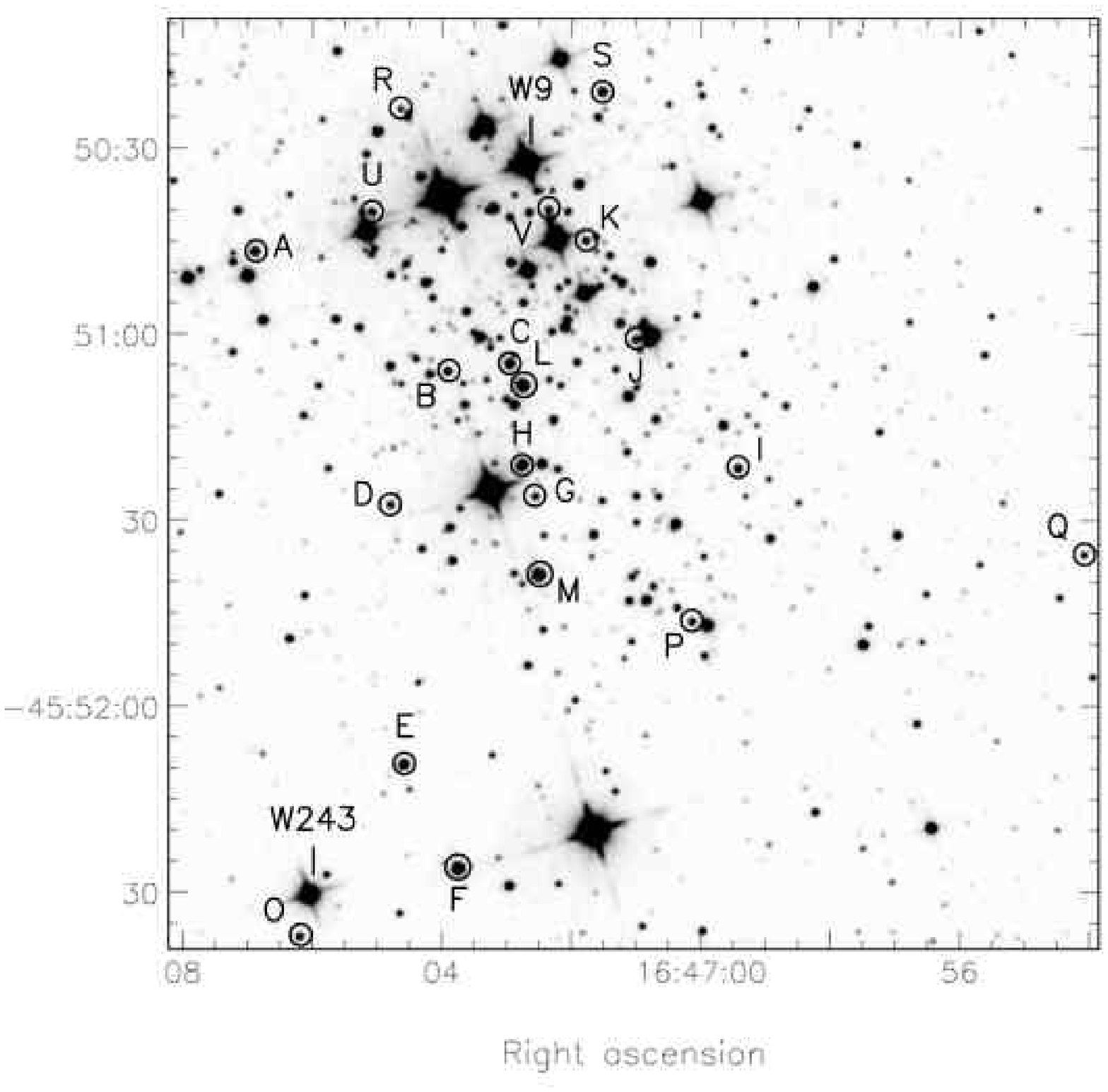,width=8.8cm,angle=0}}
\caption{(upper panel) {\it Spitzer} IRAC (3.6$\mu$m) image
of the 10$\times$10 arcmin (15 $\times 15$ pc at a distance of 
5~kpc) region surrounding Westerlund~1 surveyed with NTT/SOFI, taken
from the GLIMPSE survey (Benjamin et al. 2003).
Identifications of the four WR stars located beyond the cluster
centre (shown as a box); (lower panel) ESO/NTT SOFI Br$\gamma$ image of 
the central  2.5$\times$2.5 arcmin (3.75$\times$3.75 pc)  of  Westerlund~1, for 
which the  Wolf-Rayet stars have been indicated, 
plus W9 (sgB[e]) and W243 (LBV, Clark et al. 2005).
North  is up, with east to the left in both images.}
\label{Brg}
\end{figure}

The present study examines the Wolf-Rayet content of Westerlund~1 based
upon near-infrared narrow-band imaging and follow up spectroscopy.
High interstellar extinction ($A_{\rm  V}$= 11.6 mag, Clark et al. 2005) 
prohibits the standard optical approach of using a narrow-band He\,{\sc 
ii} $\lambda$4686 filter and adjacent continuum filter to identify WR 
candidates (e.g. Hadfield et al. 2005). Homeier et al. (2003ab) have 
previously used interference filters within the K-band to identify WR
candidates within the inner Milky Way. However, in contrast to the optical
technique, the K-band is observationally more challenging since (i) WR 
line fluxes are much weaker; (ii) there is no single pair of interference 
filters that can applied to identify all WR subtypes; (iii) dust emission
in WC stars may heavily dilute the WR stellar emission line fluxes.

The present paper is structured as follows. New K-band and Y-band imaging
of Westerlund~1 is presented in Sect.~\ref{obs} together with  
near-IR spectroscopy. Near-IR spectral classification of WR stars in 
Westerlund~1 is presented in Sect.~\ref{class}. 
The reddening and distance to  Westerlund~1 
are reassessed from its WR population in Sect.~\ref{Wd1},
and comparisons made with other Milky Way clusters hosting WR stars.
The observed WR content of Westerlund~1 is compared to predictions
from single and binary models in Sect~\ref{evol}. Finally, brief
conclusions are drawn in Sect.~\ref{conc}

\section{Near-IR observations of Westerlund~1}\label{obs}

Here we present new near-IR imaging and spectroscopy of 
Westerlund~1, obtained with the 3.5m New Technology Telescope (NTT),
La Silla, Chile, using the "Son of Isaac" (SOFI) instrument.

\subsection{Narrow-band Imaging}

A series of NTT/SOFI narrow-band images were obtained for the 
central (4.9 $\times 4.9$ arcmin) region of Westerlund~1 on 1 May 2004
(ESO programme 073.D-0321(C))
using the standard plate scale of 0.288 arcsec/pix and Hawaii HgCdTe
1024 $\times$ 1024 array. The interference filters used were
the 2.07$\mu$m (He\,{\sc i}), 2.09$\mu$m (C\,{\sc iv}), 2.13$\mu$m 
(continuum),
2.17$\mu$m  (Br$\gamma$), 2.19$\mu$m (He\,{\sc ii}) and 2.25$\mu$m 
(continuum),
each with FWHM = 0.02 -- 0.03$\mu$m. The jittered images for Wd~1 and 
Hipparcos standard stars were reduced using {\sc orac-dr} (Economou et al. 
2004). Our Br$\gamma$-band image of the central cluster (FWHM $\sim$ 0.7 
arcsec) is presented in the lower panel of Fig.~\ref{Brg}, together with 
WR stars indicated, plus W9 (sgB[e]) and W243 (LBV, Clark et al. 2005).
Four WR stars lie beyond the central  region,  which are indicated in 
the 10$\times$10 arcmin {\it Spitzer}  IRAC (3.6$\mu$m) image 
shown in the upper panel of Fig.~\ref{Brg}, taken from the GLIMPSE survey
(Benjamin et al.  2003).

Images were prepared following standard procedures, with
photometry obtained using {\sc daophot} within {\sc iraf}, with typical 
uncertainties of $\pm$0.1 mag.

In principle, different pairs of filters should permit the identification
of different WN and WC subtypes.  
%
%
Late type WN and WC
subtypes, with strong He\,{\sc i} 2.058$\mu$m emission
 (Crowther \& Smith  1996; Eenens et al. 1991),  are 
anticipated to be  brighter in the 2.07$\mu$m filter than at 2.09$\mu$m.
Early and mid-WN 
stars, with strong He\,{\sc ii} 2.189$\mu$m emission 
(Crowther \& Smith 1996) should be brighter in the
2.19$\mu$m filter than at 2.25$\mu$m. Finally, early and mid-WC stars 
should be 
much brighter
in the 2.07$\mu$m filter than at 2.12$\mu$m due to very strong C\,{\sc iv} 
2.08$\mu$m emission (Eenens et al. 1991). Complications will arise 
in case of dust emission, leading to a
dilution of emission line strengths, or extremely high interstellar
extinction, leading to differences between the continua across the K-band.

\begin{figure}
\centerline{\psfig{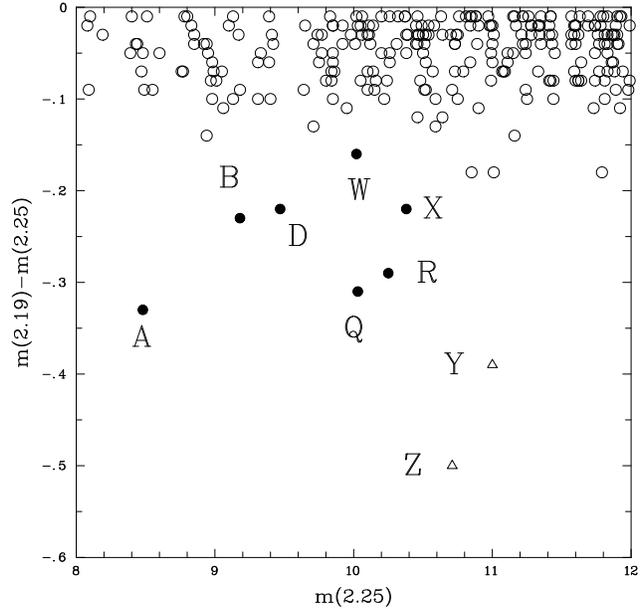}}
\caption{Comparison between SOFI interference filter photometry 
of the central cluster at 2.189$\mu$m (He\,{\sc ii}) and 2.25$\mu$m 
(continuum).  The 2.25$\mu$m zero point has been defined using K$_{S}$-band 
magnitudes for our WR stars. Strong He\,{\sc ii} emission is observed for
known early-type WN stars  within  Westerlund~1 (filled circles), including our 
newly identified
candidates W and X, Y and Z, for which the latter two were not spectroscopically
confirmed (open triangles, see text).}\label{219-225}
\end{figure}

From our anticipated combinations, the most useful proved to be the
[2.19] -- [2.25] pair, for which several known mid-type WN stars exhibited 
large $\sim$0.3 mag excesses in the 2.19$\mu$m filter. Four additional
candidates revealed an excess of at least 0.15 mag, which we have labelled 
as W, X, Y, Z, 
following our
previous nomenclature for WR stars in Westerlund~1, which ended at S (Negueruela
\& Clark 2005), supplemented by WR77aa
from Hopewell et al. (2005),  which we shall denote as source T 
hereafter plus
U and V previously discovered from optical spectroscopy by Negueruela 
(2005, priv. comm.). 

The [2.19] -- [2.25] colour magnitude diagram 
is presented in Figure~\ref{219-225}.  We set the zero point of the 
2.25$\mu$m filter to the K$_{S}$-band magnitude for simplicity 
since no flux standards were taken with this filter set. The Figure 
illustrates
how the known WN stars in Westerlund~1, A, B, D, Q and R, display the 
expected  He\,{\sc ii} 2.189$\mu$m excess. Following the completion of our 
study, 
Groh et al. (2006) presented their own Y-band imaging of Westerlund~1
which highlighted sources W (their \#3) and X (their \#2), plus
follow-up spectroscopy that included source U (their 
\#1).
%
%

The other combinations
revealed only one suitable match between them, i.e. star E (WC9) 
which showed an excess in the [2.07] -- [2.09] diagram.
Other late WN or WC stars previously known in Westerlund~1 were not 
identified from these images, although some of the brighter stars were 
saturated. Our spectroscopy (Sect~\ref{class}) revealed that
line dilution by dust emission was responsible for the lack of [2.07] -- 
[2.09] excess in several WR stars. Finally, no evidence for early WC stars 
was found from 
a comparison between the 2.07$\mu$m and 2.12$\mu$m images. 
%
%
Either early type WC stars are absent in Westerlund~1 or dust dilution
would need to be very severe to prevent their identification from narrow-band
imaging. For example, 
K-band emission lines are seen during dust formation 
in the case of WR140  (WC7+O4--5,Williams 2002).

Two WR stars associated with Westerlund~1  
are known to lie at a distance of several arcmin from the cluster core
(Fig.~\ref{Brg}). Consequently, on 29 Jun 2005,   the same SOFI K-band 
interference filters  described above were used to image fields centred 
$\sim$4  arcmin to the NE,  NW, SW and SE, i.e. sampling the 
10$\times$10 arcmin region shown in the upper panel of Fig.~\ref{Brg}.

\begin{table*}
\caption{{\it Spitzer} IRAC point source catalogue photometry from the 
GLIMPSE survey (Benjamin  et al. 2003) and mid-IR colours of selected 
Wolf-Rayet stars in Westerlund~1, plus red sources within our observed
fields exhibiting [1.08] -- [1.06] excesses of 0.4 mag or greater 
(their 2MASS designations are indicated, Skrutskie et al. 2006).}
\label{IRAC}
\begin{tabular}{l
@{\hspace{-9mm}}l
@{\hspace{2mm}}c
@{\hspace{2mm}}r
@{\hspace{3mm}}r
@{\hspace{3mm}}r
@{\hspace{3mm}}c
@{\hspace{3mm}}c
@{\hspace{3mm}}c
@{\hspace{3mm}}c
@{\hspace{3mm}}c
@{\hspace{2mm}}c
@{\hspace{2mm}}c
@{\hspace{2mm}}c}
\hline
Source & Sp Type & [1.08] -- [1.06]  & J & J--H     & H--K$_{S}$    & Ref  & 
[3.6]    & [4.5] & [5.8]    & [8.0]  &[3.6]--[4.5] & [5.8]--[8.0] & K$_{S}$--[8.0] \\
     
\hline
D             & WN7o    &--1.0 & 11.63 & 1.32     & 0.70    & SOFI & 8.47   & 8.17 & --    &       & 0.30    & --      &  -- \\
E             & WC9     &--1.0 & 10.12 & 1.03     & 0.80    & SOFI & 7.13   & 6.71 & --    & 6.14  & 0.42    & --      & 2.15 \\
I             & WN8o    &--1.4 & 10.89 & 1.32     & 0.71    & SOFI & --     & --   & 7.32  & 6.96  & --      & 0.36    & 1.90 \\
N             & WC9d    &      &  9.69 & 1.85     & 1.43    & 2MASS& --     & --   & 4.32  & 4.17  & --      & 0.15    & 2.24 \\
T             & WC9d    &--1.1:& 10.04 & 1.83     & 1.49    & 2MASS& --     & --   & 4.04  & 4.05  & --      & 0.01    & 2.67 \\
W             & WN6h    &--0.8 & 12.11 & 1.36     & 0.71    & SOFI & 9.01   & 8.61 & 8.37  & 8.01  & 0.40    & 0.36    & 2.03 \\
X             & WN5o    &--0.5 & 12.36 & 1.28     & 0.83    & 2MASS& 9.32   & --   & 8.71  & 8.35  & --      & 0.36    & 1.90 \\
16471179-4548361&       &--0.4 & 11.00 & 2.62     & 1.27    & 2MASS& --     & --   & 6.10  & 6.05  & --      & 0.05    & 1.05 \\
16464503-4548311&       &--1.5 & 11.26 & 2.53     & 1.34    & 2MASS& --     & --   & 5.85  & 5.52  & --      & 0.33    & 1.87 \\
16464447-4550044&       &--0.6 & 11.42 & 2.76     & 1.57    & 2MASS& --     & --   & 5.77  & 5.47  & --      & 0.30    & 1.62 \\
\hline
\end{tabular}
\end{table*}

In addition, due to the potential diluting influence of a 
strong dust continuum to the emission lines within the K-band, 
we obtained images for 
all pointings using two further Y-band interference filters on 30 Jun 
2005,  namely 1.06$\mu$m (continuum) and 1.08$\mu$m (He\,{\sc i}). The 
[1.08] -- 
[1.06]  index suffers from a higher sensitivity to extinction with respect 
to  the K-band indices, yet one expects He\,{\sc i} 1.083$\mu$m to be 
exceptionally 
strong in late-type WR stars of both sequences, be present in 
early-type WR stars, and not suffer the same degree of dust continuum 
dilution as those obtained in the K-band  (e.g. Howarth \& Schmutz 1992). 

\begin{figure}
\centerline{\psfig{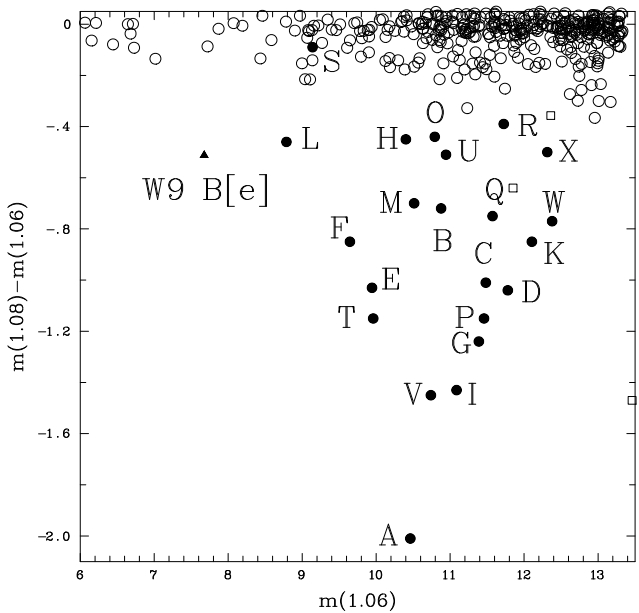}}
\caption{Comparison between SOFI interference filter photometry 
of the central cluster at 1.08$\mu$m (He\,{\sc i}) and 1.06$\mu$m 
(continuum),
illustrating
strong emission for known WN and WC stars (filled circles) plus the B[e] supergiant
W9 (filled triangle). 
The 1.06$\mu$m zero point has been defined using J-band magnitudes for 
a subset of our WR stars, 
permitting the inclusion of source T and
three extremely red sources (open squares) from narrow-band imaging in the NW and NE fields.}
\label{108-106}
\end{figure}

We present the corresponding 
[1.08] -- [1.06] diagram for the entire Westerlund~1 region in Fig.~\ref{108-106}.
We set the zero point of the 1.06$\mu$m filter to the J-band magnitude for simplicity 
since no flux standards were taken with this filter set. 
We estimate typical 
uncertainties of $\pm$0.1 mag, from comparison between photometry of individual
sources observed in multiple fields. Fig~\ref{108-106}
illustrates the much cleaner identification of WR stars
in  Westerlund~1, with the exception of source S which possesses unusually
weak, narrow He\,{\sc i} emission. The extreme B[e] supergiant W9  
(Clark et  al. 2005) is also a strong He\,{\sc i} 1.083$\mu$m 
emitter. All known and candidate WR stars are indicated here except for N which
lies at the edge of the SE field, and J which has a very bright close neighbour
preventing reliable photometry.
A similar comparison has recently been made by Groh et al. (2006) for the central
cluster, who also exploited filters centred upon He\,{\sc ii} 1.01$\mu$m 
and the adjacent 
continuum.
Of course, extremely red sources will also exhibit a large [1.08] -- [1.06] excess.
Those with excesses greater than 0.4 mag are indicated in the figure, and
will be considered in the next section.

\begin{table*}
\caption{NTT/SOFI emission equivalent widths (in \AA, accurate to $\pm$10\%) 
for prominent near-IR lines 
(wavelengths are in $\mu$m) in WN stars in Westerlund~1.
}
\label{EW-WN}
\begin{tabular}{
l@{\hspace{2mm}}
l@{\hspace{2mm}}
l@{\hspace{2mm}}
l@{\hspace{2mm}}
l@{\hspace{2mm}}
l@{\hspace{2mm}}
l@{\hspace{2mm}}
l@{\hspace{2mm}}
l@{\hspace{2mm}}
l@{\hspace{2mm}}
l@{\hspace{2mm}}
l@{\hspace{2mm}}
l@{\hspace{2mm}}
l@{\hspace{2mm}}
l@{\hspace{2mm}}
c@{\hspace{2mm}}
c@{\hspace{2mm}}
}
\hline
Star &Sp   & He\,{\sc ii} & He\,{\sc i}   & He\,{\sc ii}+P$\gamma$ & 
He\,{\sc ii}
&He\,{\sc ii}+P$\beta$& He\,{\sc ii}  & He\,{\sc ii}
& He\,{\sc i}   & He\,{\sc i}   & N\,{\sc v}    & He\,{\sc i}/N\,{\sc iii}  
& He\,{\sc ii}+Br$\gamma$ & He\,{\sc ii} & 1.012& 2.189\\
      &Type& 1.012& 1.083 & 1.093     & 1.163  & 1.281       & 1.476 &  1.692 & 1.700 & 2.058 & 2.110 & 2.115     & 2.165           & 2.189& /1.083& /2.165\\
\hline
A& WN7b    & 280  & \multicolumn{2}{c}{--- 930 ---} 
                                      & 194   & 110          & 113    & 
\multicolumn{2}{c}{--- 141 ---}  & 35   & \multicolumn{2}{c}{--- 74 ---} 
                                                                                                                  & 87         &  123 & 0.3 & 1.4 \\
B & WN7o   & 65   & 147    & 12       & 50    & 30           & 28     & 16:    & 21:   & $\sim$5& 3  & 25         & 34          &  36 & 0.4 & 1.1 \\
D& WN7o  & 98   & 316    & 21       & 88    & 47           & 39     &  18    & 34    & 13   & 1    & 35         & 68         &   47 & 0.3 & 0.7 \\
G & WN7o   & 148  & 443    & 29       & 87    & 58           & 41     & 11     & 34    & 8    & 3    & 34         & 53          & 33   & 0.3 & 0.6 \\
I & WN8o   & 54   & 459    & 32       & 44    & 61           & 22     & 7      & 48    & 26   & 3    & 50         & 64        & 17    & 0.12 & 0.26 \\
J & WN5h   & 56   & 31     & 8         & 38   & 15           & 22     &15    & $<$4& $<$4 & 3 & 12        & 31       & 46  & 1.8 & 
1.5\\
L & WN9h:  & 3    & 72     &  9       & $<$1  & 22           & $<$1   & $<$1   & 11    & 50   & 0   &  7         & 23         &  
$<$2& 0.04  & $<$0.08 \\
O & WN6o   & 93   & 82     & 7        & 66    &  21          & 37     & 18     & 8     & 4    & 3    & 17         & 20         & 44    & 1.1 & 2.2 \\
P & WN7o & 78    & 285   & 22       & 75    & 41           & 35     & 9      & 23    & 10   & 4    & 34         & 38         & 27    & 0.3 & 0.7\\
Q& WN6o    & 171  & 159    & 19       & 128   & 43           & 68     &  29    & 12    & 6    & 6    & 27         & 31         &   71 & 1.1 & 2.3 \\
R & WN5o   & 168  & 82     & 12       & 109   & 35           & 63     & 31     & 7     & 3    & 8    & 23         & 32         & 94   & 2.0 & 2.9 \\
S & WN10-11h:& $<$2 & 14    & 1       & $<$1  & 8            & $<$1   & $<$1   & 1      &  10  &   0   &  1          & 8        & $<$0.5& $<$0.07 
& $<0.06$ \\
  & or B0--1Ia$^{+}$ \\
U &  WN6o  & 56 & 72     & 6        & 42    & 14           & 27     & 14      & 7    & 4    & 2    & 15          & 17        & 34   & 0.8 & 2.0 \\   
V& WN8o    & 41   & 326    & 17       & 31    & 43           & 21     &  5     & 39    & 22   & 2    & 37         & 48         &   17 & 0.13 & 0.35 \\
W & WN6h   & 113 & 137    & 31       & 86    & 65           & 44     & 20     & 8     & $<$4 & 4    & 19         & 55         & 51   & 0.8  & 0.9 \\
X & WN5o   & 214  & 112    & 20      & 134   & 48           & 61     & 31     & 8     & 3    & 9     & 20        & 34         & 98    & 1.9 & 2.9 \\  
\hline
\end{tabular}
\end{table*}

\begin{table*}
\caption{NTT/SOFI emission equivalent widths (in \AA,  accurate to $\pm$10\%) 
for prominent near-IR 
lines (wavelenths are in $\mu$m) in WC stars in Westerlund~1.
}
\label{EW-WC}
\begin{tabular}{
l@{\hspace{2mm}}
l@{\hspace{2mm}}
l@{\hspace{2mm}}
l@{\hspace{2mm}}
l@{\hspace{2mm}}
l@{\hspace{2mm}}
l@{\hspace{2mm}}
l@{\hspace{2mm}}
l@{\hspace{2mm}}
l@{\hspace{2mm}}
l@{\hspace{2mm}}
l@{\hspace{2mm}}
l@{\hspace{2mm}}
l@{\hspace{2mm}}
l@{\hspace{2mm}}
l@{\hspace{2mm}}
l@{\hspace{2mm}}
l@{\hspace{2mm}}
l@{\hspace{2mm}}
c@{\hspace{2mm}}
c@{\hspace{2mm}}
}
\hline
Star &Sp  & C\,{\sc iii}  & C\,{\sc ii}   & He\,{\sc ii}  & He\,{\sc i}   
& C\,{\sc iv}   & C\,{\sc iii}  & He\,{\sc i}/C\,{\sc ii} & C\,{\sc iv}  
&  C\,{\sc ii}   & He\,{\sc i}   & C\,{\sc iv}   & C\,{\sc iii}  & 
He\,{\sc ii}/C\,{\sc iii} & He\,{\sc ii} &0.971 & 2.076\\  
     &Type& 0.971 & 0.990 & 1.012 &1.083  & 1.191 & 1.198 &  1.700  & 1.74 &  1.785 & 2.058 & 2.076 & 2.110 & 2.165     & 2.189 &/0.990 & /2.110\\
\hline
C    & WC9d& 305  & 44    & 17    &191    & 23    & 87    &   15    & 13   & 44      & 44   & 22    & 38    & 11        & 11    & 7 & 0.58 \\
E    & WC9& 132  & 31    & 8     &226    & 10    & 61    &   32    & 12:  & 65:     & 153  & 30    & 54    &   30      & 16    & 4.3 & 0.55 \\
F    & WC9d& 163  & 22    & 8     &189    &  8    & 37    &    5    &  2   & 24      & 2    & $<$1  & 3     & $<$1      & $<$1  & 7.5 & $<$0.3 \\
H    & WC9d&  89  & 23    & 3     & 84    & 11    & 40    &   9   &  4 & 30   
 &  37  & 13  & 23  & 9      & 8   & 3.9  & 0.56 \\ 
K    & WC8& 509  & 36    & 35    &132    & 82    & 67    &   43  &  149& 
95   & \multicolumn{2}{c}{--- 430 ---} & 198 & 39      & 59  & 14 & 2.2\\
M    &WC9d & 269  & 53    & 14    & 154   & 11    & 56    &   3     & $<$2 & 14      & 4    & 2     & 7     & $<$1      & $<$1  & 5.1  & 0.28  \\
N    &WC9d & 169 & 20  & 11   & 81   &  13  & 23  &  5    &  6  &  
9    & 5   & 9  & 11  &  5      &  4  &  8.6 & 0.8\\
T    &WC9d   &345 & 75    & 15    & 281   & 12    & 75    &  12   &  19  
&   20  &  27  &  5  &  13 &  5      &  3 & 4.6& 0.4 \\
\hline
\end{tabular}
\end{table*}

\subsection{Broad band near-IR imaging and photometry}

In addition to our interference band imaging, we obtained jittered J, H
and K$_{S}$ band images of the central region of Westerlund~1 on 
24 Sep
2004.  These were reduced using {\sc orac-dr}, from which {\sc daophot}
photometry was obtained, with typical uncertainties of $\pm$0.05 mag.  
These were supplemented by 2MASS (Skrutskie et al. 2006) 
J, H, K$_{S}$\footnote{Note that both NTT/SOFI and 2MASS employ K$_{S}$ 
filters ($\lambda_{c} \sim 2.16\mu$m, FWHM$\sim 0.27\mu$m), in contrast 
to K$'$ ($\lambda_{c} \sim 
2.12 \mu$m)  or K ($\lambda_{c} \sim 2.20 \mu$m) filters, which are 
commonly used elsewhere.} photometry for WR stars which were located 
beyond the 
cluster core, where spatial crowding was not so significant. In
a few cases SOFI and 2MASS photometry are available - notably for
source I, agreement is within 0.01 mag, except that 2MASS indicates a 
H-band magnitude fainter by 0.09 mag. 

\begin{figure}
\centerline{\psfig{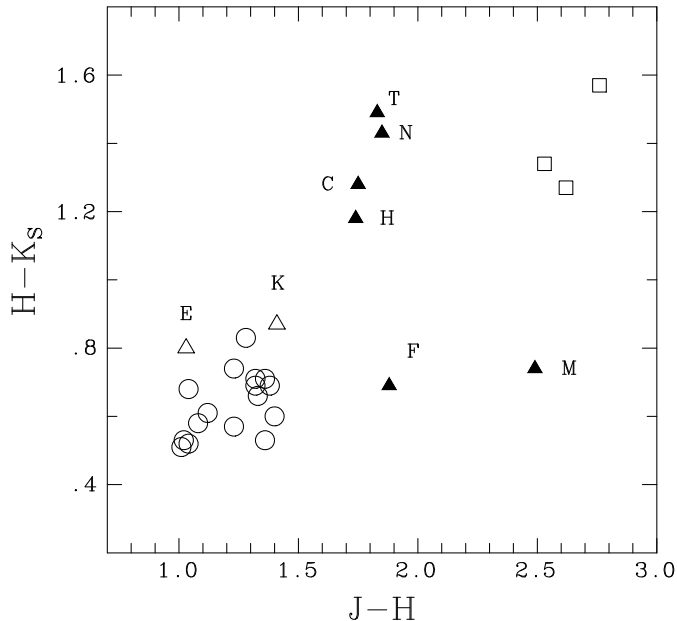}}
\caption{Near-IR colour-colour diagram for Westerlund~1 WN 
(circles) and WC (triangles) stars, plus three very red sources in
the vicinity of Westerlund~1 (squares, Table~\ref{IRAC}).
Dust-forming WC stars are indicated as filled symbols (compare their
colours to non-dusty stars E and K).}
\label{nir}
\end{figure}

A colour-colour diagram for the known WR stars in Westerlund 1 is presented 
in Figure~\ref{nir}. The scatter amongst WN stars is due to intrinsic
colour and extinction variations, whilst the WC stars are offset to 
higher values, as  a result of hot, circumstellar, dust emission, with the
exceptions of sources E (WC9) and K (WC8). We shall therefore
follow the recent convention in adding `d' to WC classifications to
signify dust emission (e.g. van der Hucht 2006). We shall address the 
presence or absence of dust in Sect.~\ref{properties}. 
We also include the three sources displaying a large [1.08] - [1.06] 
excess in Fig~\ref{nir}, which overlap with the position of the 
dust  forming WC stars. 

Further evidence for the presence of dust emission will be revealed via
line dilution in K-band spectroscopy in the next section.  
Table~\ref{IRAC} presents IRAC photometry for those WR stars in
Westerlund~1 from the GLIMPSE survey (Benjamin et al. 2003)
that are sufficiently isolated, together with mid-IR colours.  
We also include the three sources displaying a large [1.08] - [1.06]
excess.  In principle, these represent additional strong He\,{\sc i} 
emission line
candidates, although they are extremely red sources as seen from the J-H,
H-K and IRAC colours, such that we attribute the [1.08] - [1.06] excess to
their extremely red nature.  It was possible to test this suggestion
for several cases in
which SOFI spectroscopy was obtained.

As expected, the mid-IR colours for WN stars are relatively
uniform  with 
K$_{S}$ - [8.0] $\sim$ 1.95 mag, or (K$_{S}$ - [8.0])$_{0} \sim$ 
1.40 mag after correction for interstellar extinction 
($A_{\rm 8.0} \sim 
0.43  
A_{\rm K_{S}}$, Indebetouw et al. 2005). This is typical of WR stars
whose near to mid-IR spectrum is dominated by the free-free excess from
the stellar wind superimposed upon the underlying Rayleigh-Jeans tail
(e.g. Barlow et al. 1981). Non-LTE stellar atmosphere models 
(e.g. Hillier \& Miller 1998) suggest (K$_{S}$ - [8.0])$_{0}$ in the range 
from $\sim$1.4 mag 
for strong-lined WN stars such as WR40 and WR6 (Herald et al. 2001; 
Morris et al. 2004)  to $\sim$0.6 mag for weak-lined WN stars such as WR3
(e.g. Marchenko et al. 2004). The presence of hot dust in WC stars 
produces flatter mid-IR colours (e.g. Williams et al. 1987, van der 
Hucht et al. 
1996, Hopewell et al. 2005), as shown in Table~\ref{IRAC} (e.g.
compare the  [5.8] - [8.0] colours for T and N with respect to I, W and 
X.)


\subsection{Spectroscopy}

Spectroscopy of known and candidate
Wolf-Rayet stars in Westerlund~1 was obtained on 29--30 Jun 2005
with SOFI, using the IJ and HK grisms (programme 075.D-0469). A slit width 
of 0.6 arcsec provided a resolving power of $R \sim$ 1000. In general, 
position angles were 
selected in order to simultaneously observe two (or more) targets.  
Spectroscopic
datasets were reduced using {\sc iraf}, with wavelength calibration
achieved using internal arc lamps. Frequent late B-type Hipparcos
standard were obtained at an airmass close to that of Westerlund~1
in order to achieve  flux and telluric correction. Photometry from
J,H, K$_{S}$ imaging provided an absolute flux calibration.

The two candidates common to Figs.~\ref{219-225}-\ref{108-106}, namely
W and X were confirmed as Wolf-Rayet stars (see also Groh et al. 2006), whilst 
sources Y  and Z, that appeared to exhibit a strong excess at 
He\,{\sc ii} 2.19$\mu$m (but not He\,{\sc i} 1.08$\mu$m) were not. K-band
spectroscopy of Y was featureless, whilst source Z displayed 2.3$\mu$m CO 
absorption, typical of late-type giants or supergiants (its J--H colour
was also red). The origin of the apparent 2.19$\mu$m photometric excess is 
consequently unclear.

The addition of W and X brings the number of known WR stars in 
Westerlund~1 to 24, or 23 if source S is omitted due to its  
ambiguous nature. Several additional candidates identified from the extended 
field were observed  spectroscopically, including the source with 2MASS
designation 16464503-4548311
(Table~\ref{IRAC}). These sources were not confirmed as WR stars, again
revealing CO 2.3$\mu$m absorption,  such that their [1.08] -- [1.06] 
excess appears to result from intrinsic red colours in such cases.

Spectroscopy of individual WR stars is discussed in the next section,
whilst a catalogue of the known WR population in Wd~1 is presented
in Table~\ref{WRcat}. Coordinates were obtained from our 
SOFI images, except where noted, using astrometry of A--G
from 3.6~cm radio images which are nominally accurate to $\pm$0.3$''$ 
(Clark \&  Negueruela 2002). These differ
from previously published coordinates for I and J, as a result of  
severe crowding within these regions.


\section{Near-IR spectral classification of WR stars}\label{class}

Several near-IR atlases of WR spectroscopy have previously been
presented, notably by Howarth \& Schmutz (1992) in the Y-band, and
by Crowther \& Smith (1996) and Figer et al. (1997) 
for WN stars and Eenens et al. (1991) for
WC stars at longer wavelengths. However, no attempt has previously been 
made at quantifying subtypes based upon near-IR spectroscopy, with
reference to the standard optical criteria of He\,{\sc} 5876\AA/He\,{\sc 
ii} 5411\AA\ for 
WN subtypes (Smith et al. 1996) and C\,{\sc iii} 5696\AA/C\,{\sc iv} 
5805\AA\ for WC subtypes 
(Crowther et al. 1998).  In  principle, suitable classification 
diagnostics are available in the near-IR, notably He\,{\sc i} 
1.083$\mu$m/He\,{\sc ii} 1.012$\mu$m for 
helium and C\,{\sc iii} 2.110$\mu$m/C\,{\sc iv} 2.076$\mu$m for carbon. 
Additional line ratios, such
as He\,{\sc i} 2.058$\mu$m/He\,{\sc ii} 2.189$\mu$m and C\,{\sc ii} 
0.990$\mu$m/C\,{\sc iii} 0.971$\mu$m are potential 
secondary
diagnostics for late subtypes (these C\,{\sc ii} and He\,{\sc i} lines are 
very weak in
early subtypes). 
Towards this goal, we have collected  line equivalent 
widths for several WN and WC stars for which optical classifications are
well established, based primarily on near-IR spectroscopy from
Howarth \& Schmutz (1992), Eenens et al. (1991), Crowther \& Smith 
(1996) and Vacca et al. (2006). 

\begin{figure}
\centerline{\psfig{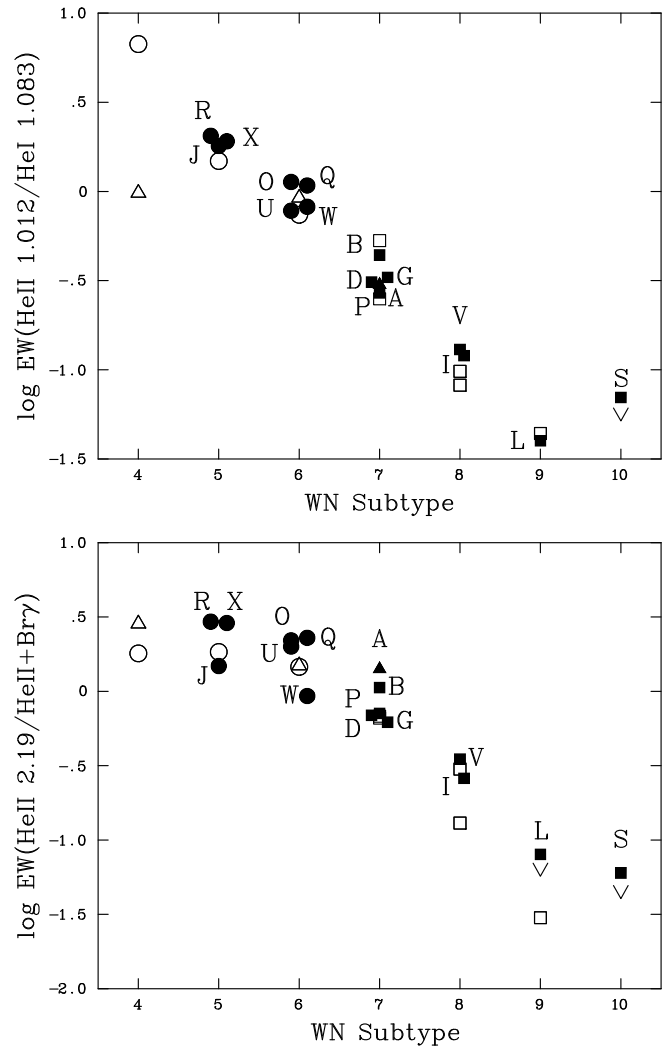}}
\caption{
Near-IR diagnostic line ratios for Westerlund 1 WN stars
(filled symbols) together with selected optically classified Milky Way stars
(WR6, 16, 78, 105, 115, 120, 123, 128, 136, 138, open symbols) 
for which
 equivalent width measurements are from Howarth \& Schmutz (1992), Crowther \& Smith 
(1996) and Vacca et al. (2006). Weak-lined WNE,
strong-lined WNE, and WNL stars are indicated by circles, triangles and squares,
respectively.}\label{WNratio}
\end{figure}

Equivalent widths of emission lines in Westerlund~1 have been measured
from normalized NTT/SOFI spectroscopy, which are presented for WN and WC
subtypes in Tables~\ref{EW-WN} and \ref{EW-WC},  respectively. Of the 
24 known WR stars
in Westerlund~1 (23 excluding source S), 16 (15) are WN-type and 8 are
WC-type, such that N(WC)/N(WN)$\sim$0.5 by number.
As such, Westerlund 1 hosts  8\% of the known Milky Way 
Wolf-Rayet population (van der Hucht 2006).

\begin{figure}
\centerline{\psfig{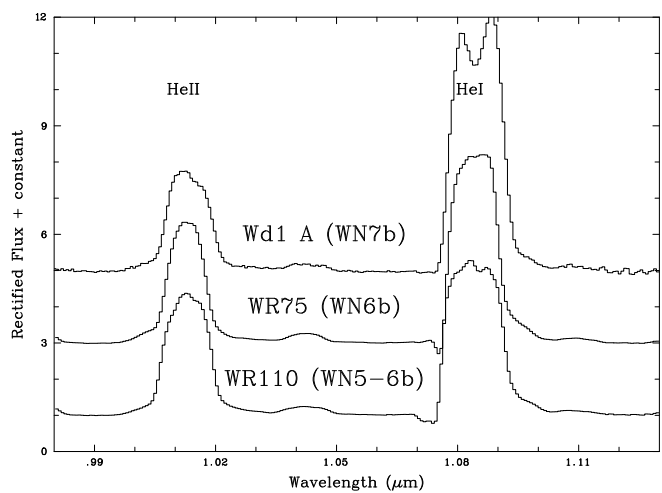}}
\caption{Spectroscopic comparison of Westerlund~1 source
A (WN7b) with other
broad-lined WN stars, WR110 (WN5-6b) and WR75 (WN6b), previously 
observed  with NTT/SOFI (Homeier, priv. comm.).}
\label{A}
\end{figure}

\subsection{WN stars}

He\,{\sc ii} 1.012$\mu$m/He\,{\sc i} 1.083$\mu$m and He\,{\sc ii} 
2.189$\mu$m/Br$\gamma$ line
ratios for selected Milky Way WN stars are presented in
Fig.~\ref{WNratio}. The Y-band ratios broadly follow optical
classification criteria, and serve as excellent discriminators between
WN4--6, WN7, WN8 and WN9 subtypes (see also Figer et al. 1997).  For
WN4--6 stars, He\,{\sc ii}/He\,{\sc i} does not in general provide a 
unique spectral type,
and instead groups stars according to strong, broad lined subtypes versus
weak, narrow-lined subtypes (Hiltner \& Schild 1966; Hamann et al. 1993).  
Smith et al. (1996) define broad lined WN stars as those for which FWHM
(He\,{\sc ii} 4686\AA)  $\geq$ 30\AA. Extending this definition to the 
near-IR, we
may identify WN stars as strong/broad if FWHM (He\,{\sc ii} 1.012$\mu$m) 
$\geq$ 65\AA\ and/or FWHM(He\,{\sc ii} 2.1885$\mu$m) $\geq$ 130\AA, for 
which solely source A qualifies from Westerlund~1.

Individual stars are now discussed in turn, starting with the latest 
subtypes.
He\,{\sc ii} 1.012$\mu$m is weak in star L, and potentially absent in S, 
indicating subtypes of WN9 and WN10-11: for these stars, 
respectively. Indeed, the weakness of He\,{\sc ii} 1.012$\mu$m in source 
S prevents
an unambiguous classification, for which an alternate early B hypergiant
classification (B0--1Ia$^{+}$) could be assigned. We favour a 
WN subtype due, in part, to the modest absolute magnitude inferred from
our adopted distance in Sect.~\ref{Wd1}.

He\,{\sc ii} 1.012$\mu$m/He\,{\sc i} 1.083$\mu$m indicators suggest WN8 
subtypes for V and I, in 
agreement with earlier far-red spectral classifications.  B, D, G and P 
have very similar near-IR line ratios, indicative of WN7 subtypes. For 
source W, the He\,{\sc ii} 2.189/Br$\gamma$ ratio suggests a subtype of 
WN7 (Figer 
et al. 1997, Groh  et al. 2006). However, the ratio of N\,{\sc 
iii}+He\,{\sc i} 2.115$\mu$m/He\,{\sc ii} 
2.189$\mu$m is 0.4  versus 0.7--1 in known Milky Way WN7 stars, such that we 
prefer WN6h (see later). Source A is unique amongst the WN stars in 
Westerlund~1 in  displaying a strong, broad lined spectrum. This star is 
reminiscent of HD~165688 (WR110, WN5--6b) and HD~147419 (WR75, WN6b)
except  for an even stronger He\,{\sc i} line strength such that we assign 
WN7b. 
1$\mu$m NTT/SOFI spectroscopy of these stars are compared in Fig.~\ref{A}, 
for which star A shows an apparent He\,{\sc i} 1.083$\mu$m absorption feature, 
suggestive of
an early B supergiant companion, which is also seen in He\,{\sc i} 
2.058$\mu$m.

\begin{figure}
\centerline{\psfig{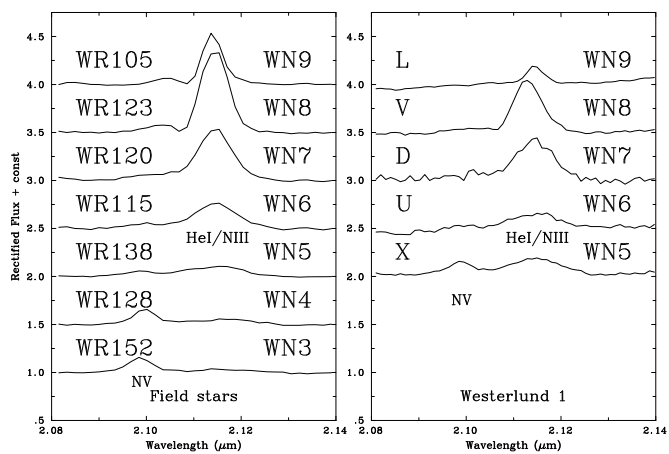}}
\caption{Spectral comparison of field weak-lined WN stars in the 
region of N\,{\sc v} 2.100$\mu$m and He\,{\sc i}/N\,{\sc iii} 2.115$\mu$m 
(from
Crowther \& Smith 1996) with selected Westerlund~1 stars}
\label{nv}
\end{figure}

WN5--6 subtypes are obtained for the other weak-lined WN stars.
For these stars, spectral types are further 
refined using their spectral morphology in the 2.11$\mu$m region, 
which is a blend of N\,{\sc v} 2.100$\mu$m and N\,{\sc iii}/He\,{\sc i} 
2.115$\mu$m.
N\,{\sc v} is relatively  strong in WN3--4 stars and N\,{\sc iii}/He\,{\sc 
i} is 
relatively 
strong in WN5--6 stars,  as illustrated in Fig.~\ref{nv}, from 
which subtypes  may be obtained, and as such this represents the best 
near-IR discriminator between such subtypes. Attempts at classification of 
broad-lined WN stars using this approach is not possible due to line 
blending.

\begin{figure*}
\centerline{\psfig{figure=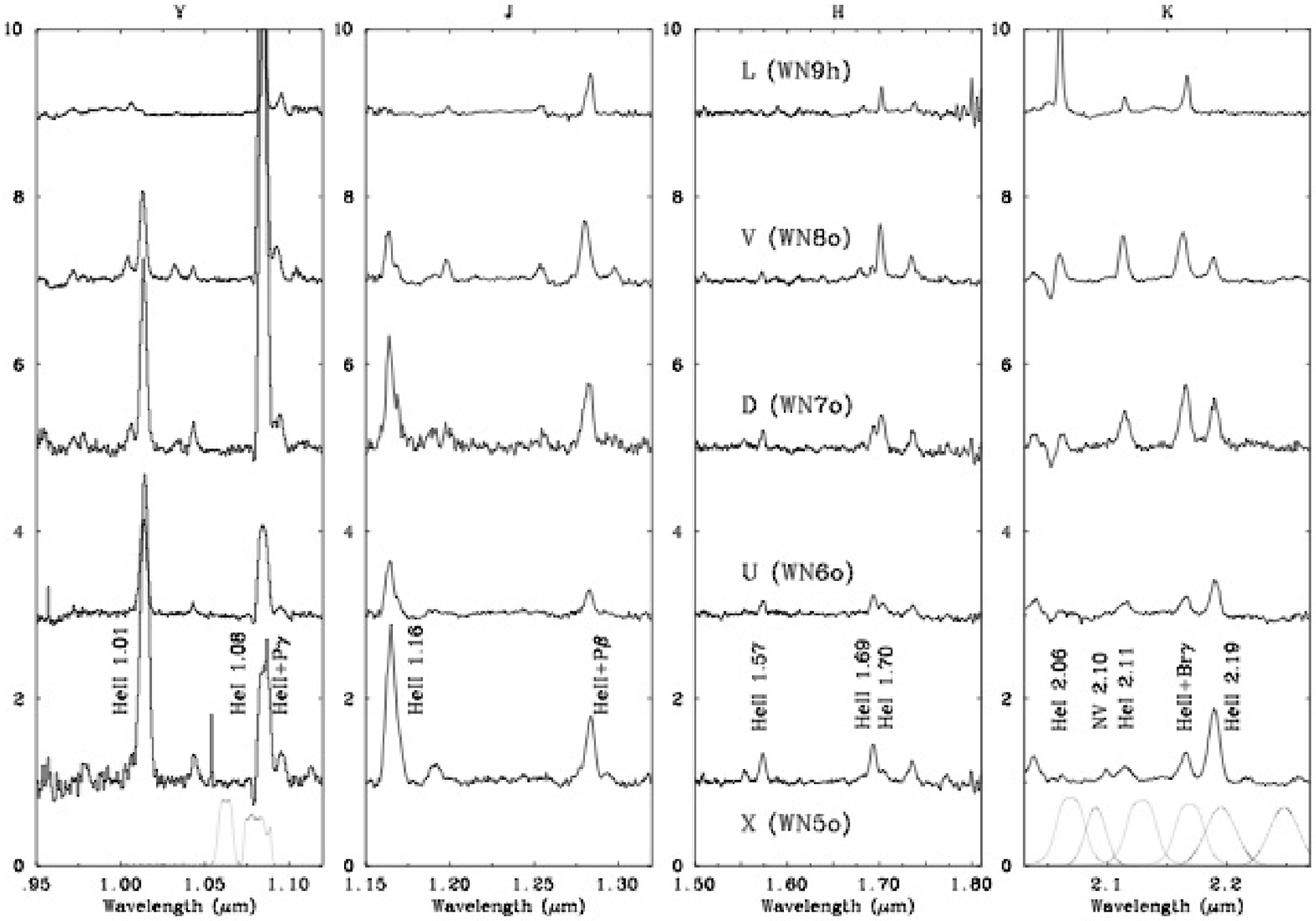,width=18cm,angle=-90}}
\caption{Near-IR spectroscopy of representative WN stars in Westerlund~1.
NTT/SOFI Y-band and K-band interference filters used in this study are 
also indicated
(dotted lines).}
\label{WNatlas}
\end{figure*}

\begin{table}
\caption{Quantitative near-IR classification of weak/narrow WN 
stars with FWHM(He\,{\sc ii} 1.012$\mu$m)$\leq$65\AA. Values for 
strong/broad lined WN
stars are shown in parenthesis.}\label{WN-summary}
\begin{tabular}{
l@{\hspace{2mm}}
c@{\hspace{2mm}}
c@{\hspace{2mm}}
c@{\hspace{2mm}}
c@{\hspace{2mm}}
c@{\hspace{2mm}}
c}
\hline
Subtype     & \multicolumn{2}{c}{He\,{\sc ii} 1.012/}  
&\multicolumn{2}{c}{N\,{\sc v} 2.100/}      & \multicolumn{2}{c}{He\,{\sc 
ii} 2.189/}\\
            & \multicolumn{2}{c}{He\,{\sc i} 1.083}   & 
\multicolumn{2}{c}{He\,{\sc i}/N\,{\sc iii} 2.115} 
&\multicolumn{2}{c}{Br$\gamma$}\\
\hline
WN3        &    $>$ 10 &         &  $>$ 2&          &  1-3 & \\
WN4        &     3-10 &($>$0.8) &    1-2 &  (blend) &  1-3 & 
($>$2)\\
WN5        &     1.5-3&         & 0.2-1&         &  1-3 &     
\\
WN6        &   0.6-1.5&($>$0.5)  &$<$ 0.25&(blend) & 0.5-2.5& 
(1-2)\\
WN7        &  0.2-0.6&($>$0.5) &        &(blend) &  0.5-1.2&(1-2) \\
WN8        & 0.07-0.2&   &         &      &  0.1-0.4 & \\
WN9        &  $<$ 0.07  &   &         &      &  $<$0.1  &\\ 
\hline
\end{tabular}
\end{table}

Historically, the optical Pickering-Balmer decrement has been used to
detect the presence of hydrogen in WN stars  (Conti et al. 1983).
Indeed, this is quantitatively included in the Smith et al. (1996) 
classification scheme, adopting o, (h), h for stars with no, weak or
strong signatures of hydrogen. The question of hydrogen in WN stars is important, 
since from an evolutionary perspective, WN stars with/without hydrogen 
are identified as late/early-types, respectively, whilst the 
spectroscopic definition is dependent upon the observed degree of 
ionization.

For the WN5--6 stars, He\,{\sc ii} 2.189$\mu$m$\gg$Br$\gamma$, suggesting
no hydrogen in most cases, i.e. WN5--6o. Stars J (WN5) and W (WN6) 
are obvious exceptions (lower panel in Fig.~\ref{WNratio}), from which we 
infer the  presence of hydrogen
in their atmospheres.  For the WN7--9 stars, the presence of hydrogen can 
in  principle be  observed in the higher Paschen (n-3) or Brackett (n-4) 
series, relative to the alternating He\,{\sc ii} (n-6, n-8) series, 
although the 
contribution of 
hydrogenic He\,{\sc i} lines to the Paschen/Brackett lines is significant 
in 
late WN stars. The observed He\,{\sc ii} 2.189$\mu$m/Br$\gamma$ ratio for 
both I and V 
mimics the H-deficient WN8 star WR123 (Crowther et al. 1995) such that
they are also expected to be strongly H-deficient. The same 
conclusions are reached for the WN7 stars in Westerlund~1 from
comparison with WR120 (WN7o, Fig.~\ref{WNratio}). 
The situation is less  clear for L and S,
but we shall assume they contain hydrogen since all other WN9--11 stars
are hydrogen rich. Of course, a detailed  determination of abundances in WN stars await 
the result of quantitative analysis, but our initial inspection suggests 
that 12/16 ($\sim$75\%) of the WN stars in Westerlund~1 are highly deficient in 
hydrogen.

Representative near-IR spectra for Westerlund~1 WN stars are 
presented in Fig.~\ref{WNatlas}, together with the Y-band and K-band 
interference filters used in Sect.~\ref{obs}. For reference,
a summary of our quantitative near-IR classification for WN 
stars is presented in Table~\ref{WN-summary}.

\subsection{WC stars}

\begin{figure}
\centerline{\psfig{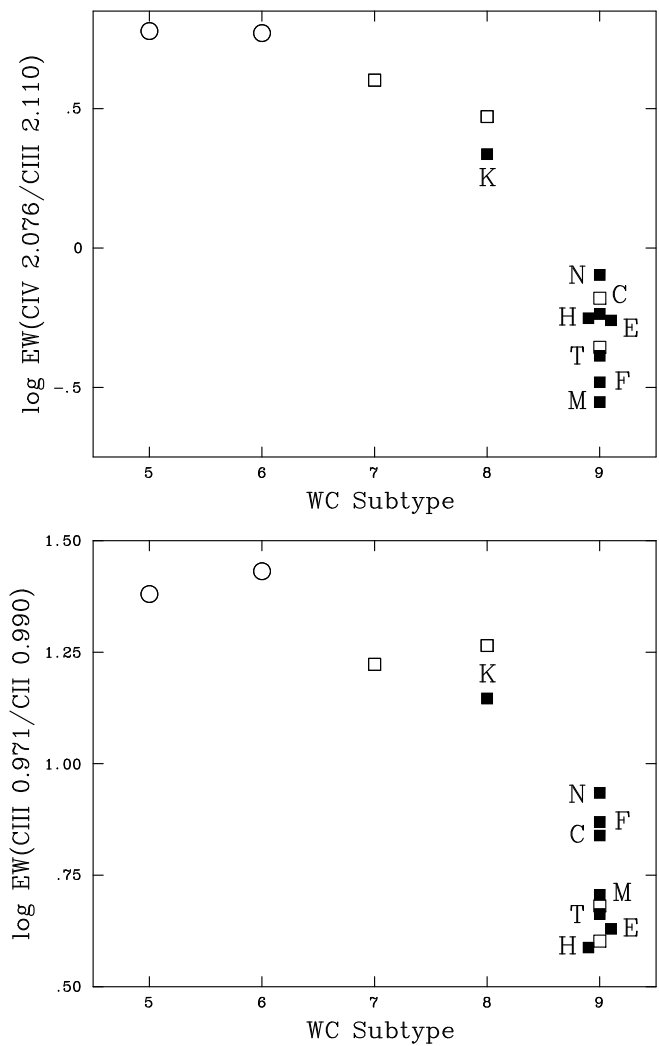}}
\caption{
Near-IR diagnostic line ratios for Westerlund 1 WC stars
(filled symbols) together with selected optically classified Milky Way stars
(WR75c, 106, 111, 121, 135, 137, 154, open symbols) for which
 equivalent width measurements are from Howarth \& Schmutz (1992), 
Eenens et al. (2001),  Vacca et al. (2006) plus our own NTT/SOFI observations
of WR75c. Early and late WC stars are indicated by circles and squares, respectively.}
\label{WCratio}
\end{figure}

The observed line ratios of C\,{\sc iv} 2.076$\mu$m/C\,{\sc iii} 
2.110$\mu$m and C\,{\sc iii} 0.971$\mu$m/C\,{\sc ii} 0.990$\mu$m
are presented for Galactic WC stars in Fig.~\ref{WCratio}. As with WN 
stars, the 
near-IR line ratios permit one to discriminate between WC5--7, WC8 and WC9 
stars. C\,{\sc ii} 0.990$\mu$m is uniformly weak in early subtypes, and 
C\,{\sc iv} 2.076$\mu$m/C\,{\sc iii}
2.110$\mu$m remains fairly constant amongst WC5--7 subtypes (Figer et al. 
1997), which is also true of other non-classification optical line ratios 
e.g. C\,{\sc iv} 5805\AA/C\,{\sc iii} 6740\AA.

\begin{figure}
\centerline{\psfig{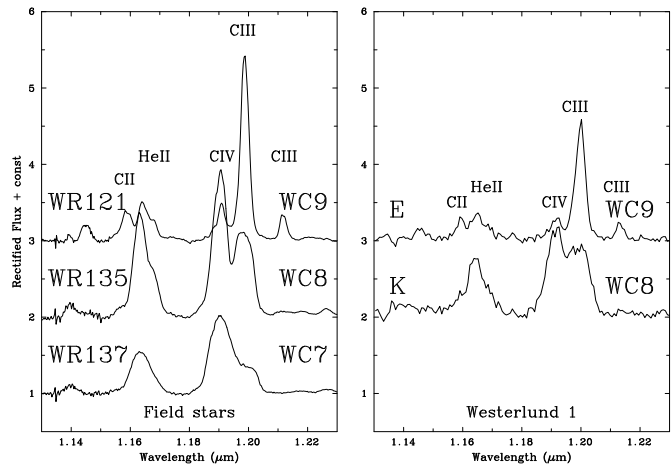}}
\caption{Spectral comparison of field WC stars in the 
region of C\,{\sc iv} 1.19$\mu$m and C\,{\sc iii} 1.20$\mu$m (Vacca et al. 
2006) with  selected Westerlund~1 stars. These lines are hard to
resolve in broad lined WC stars, typical of WC7 and earlier subtypes.}
\label{civ}
\end{figure}

The K-band C\,{\sc iv}/C\,{\sc iii} ratio serves as our primary  
classification
diagnostic, from which a subtype of WC8 is inferred for source K. 
WC9 subtypes are indicated for the remainder, with measured C\,{\sc 
iv}/C\,{\sc iii}
ratios similar to known WC9 stars, with the possible exception of star N. 
This has the highest ionization of WC9 stars in Westerlund~1 from both 
the Y-band and K-band criteria, albeit hindered by a 
heavily dust diluted K-band spectrum, so  it is necessary to 
consider whether WC8 or WC9
is most appropriate for source N, given that C\,{\sc ii}  1.78$\mu$m is 
also 
weak/absent, suggestive  of a WC8 subtype. 

Fortunately, a third near-IR diagnostic is available using the ratio 
C\,{\sc iv}
1.191$\mu$m/C\,{\sc iii} 1.198$\mu$m, from which a conclusive 
classification is
possible, as presented in Fig.~\ref{civ}. EW(C\,{\sc iv} 
1.191$\mu$m/C\,{\sc iii}
1.198$\mu$m)=1.5 for WR135 (WC8)  and only $\sim$0.2 for WR75c and WR121
(WC9).  This supports a WC9 subtype for star N since its C\,{\sc iv}
1.191$\mu$m/C\,{\sc iii} 1.198$\mu$m ratio is 0.4. Indeed, red optical
spectroscopy published by Negueruela \& Clark (2005) reveals C\,{\sc iii}
5696\AA\ $\gg$ C\,{\sc iv} 5805\AA\ from which an unambiguous
WC9 subtype follows. In addition, C\,{\sc ii} $\lambda$7240 emission is
prominent, in common with other optically visible WC9 stars.

Representative near-IR spectra for WC stars are presented in
Fig.~\ref{WCatlas}, for which three WC9 stars whose emission line
spectra are progressively diluted by dust emission are included.
Y-band and K-band interference filters used in Sect.~\ref{obs} 
are also indicated.
Our quantitative near-IR WC classification scheme is outlined in 
Table~\ref{WC-summary}.


\begin{table}
\caption{Quantitative near-IR classification of WC stars}\label{WC-summary}
\begin{tabular}{lccc}
\hline
Subtype     & C\,{\sc iii} 0.971/ & C\,{\sc iv} 1.191/ & C\,{\sc iv} 
2.076/\\
            & C\,{\sc ii} 0.990   & C\,{\sc iii} 1.198  & C\,{\sc iii} 
2.110\\
\hline
WC5--6      &  $>$15      & $>$ 4   & $>$5\\
WC7         &  $>$15      &  2-4    & $>$4\\
WC8         &  $>$10      & 0.8-2        & 1--4\\
WC9         &  $<$10      & $<$0.8       & $<$ 1 \\
\hline
\end{tabular}
\end{table}

\section{Westerlund~1 in context}\label{Wd1}

In this section we use the Wolf-Rayet population of Westerlund~1 to 
estimate its global properties, namely distance, extinction and age.
In addition, estimates of the current WR masses are presented.

\subsection{Distance and extinction to Westerlund~1}\label{distance}

We may used the Wolf-Rayet stars within Westerlund~1 to estimate its
distance and foreground extinction, independently of yellow hypergiants,
from which Clark et al. (2005) estimated $A_{\rm V}$=11.6 mag and an 
upper limit of 5.5~kpc. A revised calibration of absolute K-band 
magnitudes and  intrinsic JHK colours for WN stars and non-dusty WC stars 
is presented in Appendix~A, which is updated from van der Hucht  \& 
Williams (2006) to include theoretical energy distributions. 
Generic theoretical near-IR colours should be reliable to $\pm$0.1 mag.

Interstellar K$_{S}$-band extinctions are estimated independently
using the J--K$_{S}$ and H--K$_{S}$ colours for individual WN stars, 
 adopting the Indebetouw et al. (2005) relationship between J, H and 
K$_{S}$ band extinctions, which are fairly line-of-sight independent, 
i.e.

\[  A_{K_{S}} = 1.82^{+0.30}_{-0.23} E_{\rm H-K_{S}} \]

\noindent and

\[  A_{K_{S}} = 0.67_{-0.06}^{+0.07} E_{\rm J-K_{S}} \]

%
%

We used the average of the two  approaches for our
adopted K$_{S}$-band extinction, $\overline{A_{K_{S}}}$. 

 We obtain a mean extinction of $\overline{A_{K_{S}}}$
 = 0.96 mag for 16 WN stars, plus the non-dusty WC8--9
stars, from which mean H and J band extinctions of
1.5~mag and  2.4~mag follow, respectively,
according to Indebetouw et al. (2005). We note the
small standard deviation from which one may infer a rather uniform
extinction towards Westerlund~1. Westerlund~1 does appear to suffer from 
an anomalous visual extinction law, as discussed by Clark et al. (2005), 
so
$A_{\rm V}$ does not directly follow, without a determination of $R_{\rm
V} = A_{\rm V}/E(B-V)$. For this reason we do not attempt to derive
$A_{\rm V}$ from our near-IR observations. Nevertheless, the Galactic IR 
extinction law is independent of sight-line, such that our near-IR 
extinction estimate ought to be robust.

%

\begin{table}
\caption{Distance modulus and extinction to Westerlund~1 based upon the 
near-IR photometry of its member WN stars plus  non-dusty WC stars (E 
and K). The K$_{S}$-band extinction is estimated using
both the J--K$_{S}$ and H--K$_{S}$ colours, as discussed in the text. The average
distance modulus (DM) equates to  5.0\,kpc, in reasonable agreement 
to that obtained from yellow hypergiants by Clark et al. (2005)}
\label{DM}
\begin{tabular}{
c@{\hspace{2mm}}
l@{\hspace{3mm}}
l@{\hspace{2mm}}
l@{\hspace{3mm}}
l@{\hspace{2mm}}
r@{\hspace{3mm}}
r@{\hspace{3mm}}
r@{\hspace{3mm}}
c@{\hspace{3mm}}
l@{\hspace{3mm}}
l@{\hspace{3mm}}
l}
\hline
Star & SpType  & K$_{S}$ & A$_{\rm K_{S}}^{J-K_{S}}$ & A$_{\rm 
K_{S}}^{H-K_{S}}$ & 
$\overline{A_{\rm K_{S}}}$ & K$_{S}$ - $\overline{A_{\rm K}}$ &DM \\
\hline 
A    & WN7b & 8.37 & 0.89  & 0.85 & 0.87 & 7.50 & 12.27 \\
B    & WN7o & 9.18 & 0.99  & 0.91 & 0.95 & 8.23 & 14.15 \\
D    & WN7o & 9.61 & 1.19  & 1.07 & 1.13 & 8.48 & 14.40 \\
G    & WN7o & 9.28 & 1.23  & 1.05 & 1.14 & 8.14 & 14.06 \\
I    & WN8o & 8.86 & 1.19 &  1.09 & 1.14 & 7.72 & 13.64 \\
J    & WN5h & 9.70 & 1.11 & 0.80 & 0.95 & 8.75 & 13.15 \\
L    & WN9h:& 7.19 & 1.10 & 0.76 & 0.93 & 6.26 & 12.18 \\
O    & WN6o & 9.45 & 0.81 & 0.67 & 0.74 & 8.71 & 13.12 \\
P    & WN7o & 9.26 & 1.04 & 0.84 & 0.94 & 8.32 & 14.25 \\
Q    & WN6o &10.00 & 0.92 & 0.93 & 0.92 & 9.08 & 13.48 \\
R    & WN5o &10.26 & 0.87 & 0.76 & 0.82 & 9.44 & 13.85 \\
S& WN10-11h & 8.29 & 0.81 & 0.71  & 0.76 & 7.53 & 14.40 \\
U &   WN6o & 9.20 & 0.82 & 0.65 & 0.74 & 8.46 & 12.87 \\
V &   WN8o & 8.76 & 1.17 & 1.00 & 1.08 & 7.68 & 13.60\\
W &  WN6h & 10.04 & 1.15& 1.00 & 1.08 & 8.96 & 13.37 \\
X & WN5o & 10.25 & 1.18 & 1.22 & 1.20 & 9.05 & 13.46 \\
\\
K & WC8  & 9.53  & 0.98 & 0.89  & 0.94 & 8.59 & 13.24 \\
E & WC9  & 8.29  & 0.89 & 0.98  & 0.94 & 7.35 &       \\
\hline
     &     &       &      &      &  0.96 &     &  13.50 \\
     &     &       &      &      & $\pm$0.14 && $\pm$0.66 \\
\hline
\end{tabular}
\end{table}

From our study, the mean distance modulus is 13.50~mag, corresponding
to a distance of $\sim$5.0 kpc,  somewhat higher than
Brandner et al. (2006) who obtained 13.0$\pm$0.2 ($\sim$4.0 kpc). 
Indeed we wish to emphasise that Wolf-Rayet stars do not represent 
ideal distance calibrators, owing to the substantial scatter in their
absolute magnitudes within individual subtypes. For example, 
--4.9 $\leq$ M$_{K_{S}} \leq$ --6.8  for WN7--9 
stars (Table~A1).
This scatter is the likely origin of the  observed dispersion of $\sigma$=0.7 
in the distance modulus for Westerlund~1. In addition,
we shall show in Section~\ref{properties} that several WN stars are
thought to be binaries, affecting their role as distance calibrators.
Fortunately, the contribution from an OB companion affects the K-band 
magnitude of the Wolf-Rayet star  much less severely than in the V-band, 
since WR stars possess relatively flat spectral energy distributions with 
respect to O stars, due to the free-free continuum emission from their 
winds. 

The distance to Westerlund~1 may confidently be assumed to be 
$5^{+0.5}_{-1.0}$ kpc
unless binarity systematically affects the present results.
For an assumed Solar galactocentric distance
of 8.5~kpc, this places Westerlund~1 at a galactocentric distance 
of  4.2$^{+0.75}_{-0.35}$ kpc, i.e. similar to 
the outer edge of the  Galactic bar, which also extends to 4.4 $\pm$ 0.4 kpc 
(Benjamin et al. 2005). This location may prove significant for the
formation of a 10$^{5} M_{\odot}$ cluster within the Milky Way.

The Galactic oxygen metallicity gradient has recently been estimated from
optical recombination lines by Esteban et al. (2005) revealing
$\Delta \log$(O/H) = --0.044 $\pm$ 0.010 dex kpc$^{-1}$, such
that the metallicity of Westerlund~1 is anticipated to be moderately
oxygen-rich, exceeding that of the Orion Nebula by 60\%, i.e. 
$\log$ (O/H) + 12 $\sim$  8.87.

\begin{figure*}
\centerline{\psfig{figure=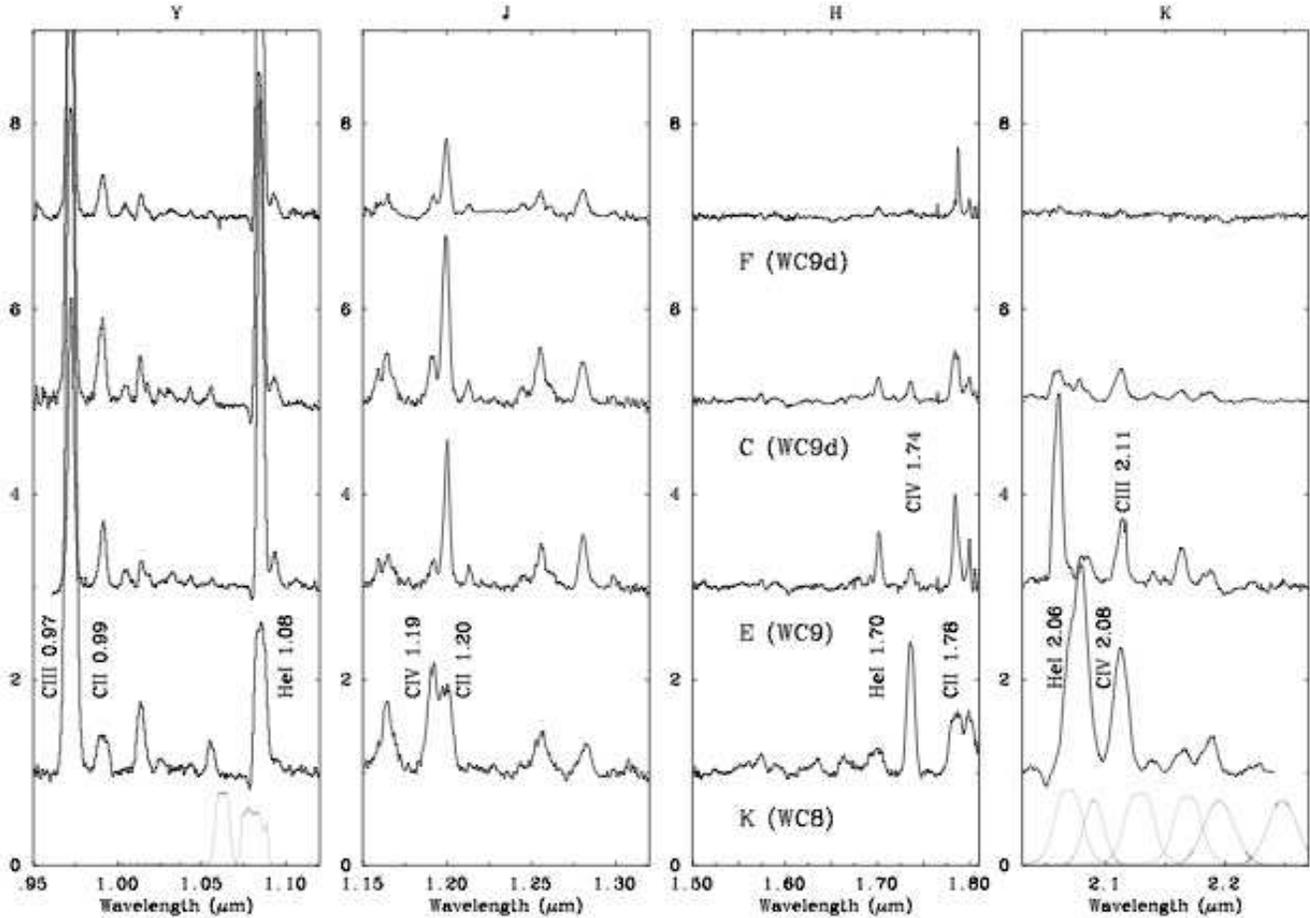,width=18cm,angle=-90}}
\caption{Near-IR spectroscopy of representative WC stars in Westerlund~1.
NTT/SOFI Y-band and K-band interference filters used in this study are 
also indicated
(dotted lines).}
\label{WCatlas}
\end{figure*}

\subsection{Age}

Clark et al. (2005) estimated a mass of 10$^{5} M_{\odot}$ for
Westerlund~1 on the basis of an age of 4--5 Myr and a Kroupa initial
mass function (IMF). This choice of IMF appears to be borne out
by recent VLT/NACO imaging (Brandner et al. 2006), whilst the 
age was selected on the basis of the simultaneous presence of 
WR stars, red supergiants and yellow hypergiants within the cluster.
Red supergiants are predicted after 4~Myr, whilst WR populations
decline rapidly after 5~Myr, especially those based on single star
evolutionary models. Definitive near-IR studies of the main sequence 
stars in Westerlund~1 are ongoing, although late O stars 
appear to be present, 
such that the turn-off age is in broad agreement with 4--5Myr. 

The ratio of the red supergiants and yellow hypergiants to the number of 
WR stars in Westerlund 1, N(RSG + YHG)/N(WR) = 8/24 = 1/3, 
including the recently identified red supergiant W75 (Clark, priv. comm.). 
In principle, this should
provide robust constraints upon its age. Vanbeveren et al. (1998)
show the lifetimes of these phases versus initial mass for single stars,
suggesting a negligible red supergiant lifetime for initial masses in 
excess of 40 $M_{\odot}$ and a decreasing WR lifetime below this initial
mass. According to Eldridge \& Tout (2004) models, the observed ratio
is best reproduced at an age of 4.5--5.0 Myr. As we shall see,
stellar evolution models poorly predict individual WR subtype 
distributions. Nevertheless, predicted total WR populations are rather 
well matched, validating our present approach.

\subsection{WR stars in Milky Way clusters}

From membership of WR stars in open clusters, Schild \& Maeder (1984) and
Massey et al. (2001) investigated the initial masses of WR stars
empirically, from the membership of WR stars in Milky Way clusters, to
which we can now add Westerlund~1, which hosts more WR stars than all the
other optically visible cluster combined. A compilation of
optically visible Milky Way clusters containing WR stars is shown in
Table~\ref{cluster}, where we have derived the highest (initial) mass
OB star for each cluster following the methodology of Massey et 
al. (2001), except that we take account of the revised spectral 
type-$T_{\rm eff}$ calibration of Martins, Schaerer \& Hillier (2005)
for O stars, Crowther, Lennon \& Walborn (2006) for B supergiants,
plus the high mass-loss Meynet et al. (1994) Solar metallicity isochrones.
Highly reddened clusters, such as the 
Quintuplet cluster at the Galactic Centre  and SGR 1806-20 are also omitted. 
To illustrate our approach, we list parameters for the highest 
initial mass OB stars within each cluster in Appendix~B.

Hydrogen-rich WN stars are observed in 
young massive clusters, indicating initial masses in excess of $\geq 
60-110 M_{\odot}$), where the effect of uncertainties in 
distance moduli are shown for clusters within the Car~OB1
association (Humphreys 1978; Massey \& Johnson 1993). 
Indeed, these are believed to 
be core-H  burning massive O stars with strong winds (Langer et al. 1994;
Crowther et al. 1995). 
Lower mass progenitors of $\geq$40$M_{\odot}$ are 
indicated for WC9  stars together with  mostly  weak-lined 
hydrogen free WN5--6 stars based primarily upon their large population in 
Westerlund~1, plus Sand~5 (WO) in Berkeley~87.
Some early WN progenitors appear to originate 
from lower mass stars, suggesting a low mass cutoff to the
formation of a WR star no higher than 20--25$M_{\odot}$ at Solar metallicity.

\begin{table*}
\caption{Optically visible Milky Way 
clusters hosting WR stars, sorted in order of decreasing initial mass of 
the most massive OB star according to Meynet et al.  (1994) Solar  
metallicity isochrones. Our approach updates that of from Schild \& Maeder 
(1984) and Massey, DeGioia-Eastwood \& Waterhouse (2001) to take 
account of the revised OB temperature calibrations of 
Martins et al. (2005) and Crowther et al. (2006). We include an estimate 
for the cluster turn-off mass for Westerlund~1}\label{cluster} 
\begin{center}
\begin{tabular}{l@{\hspace{2mm}}clccl@{\hspace{2mm}}ll}
\hline
Cluster  & DM   & Ref & Age     &  Mass &  WR &Star &Subtype   \\
         &(mag) &    & (Myr)   & ($M_{\odot}$) & Cat & &      \\
\hline
Cyg~OB2   &11.2 &  5 &2$\pm$0.2  &110:& 
WR144& & WC4\\
NGC~3603        &14.25  &2  & 1.3$\pm$0.3 & 105 & WR43a,b,c& 
HD~97950A1,B,C &WN5ha\\
Pismis 24       & 12.0 & 8 & 1.8$\pm$0.2& 105 & WR93 & HD~157504 & 
WC7\\ 
Trumpler~16     &12.1,12.5&1,6&1.5$\pm$0.5&80,105 & WR25& 
HD~93162&WN6ha\\ 
Havlen-Moffat 1 &12.0&8 &2.2$\pm$0.5 &75& WR87& 
LSS~4064&WN7h\\
                &      &   &     &      & WR89& LSS~4065  &WN8h\\
NGC~6231        & 11.5 &  3  & 2.7$\pm$0.5 & 75 & WR78& HD~151932 & 
WN7h  \\
Collinder~228   
                &12.1,12.5& 1,6&2.5$\pm$0.5 &65,100 & WR24 & HD~93131 
& 
WN6ha \\
Trumpler~27     &12.0 & 8  & 3.5$\pm$1.0 & 50 &WR95&He3~1434&WC9 
\\
Westerlund~2    & 14.5 & 4& 2.6$\pm$0.2 & 50  & WR20a &  & WN6ha+WN6ha\\
Berkeley~87     & 11.0 & 8 & 3$\pm$1.0 &50  & WR142& Sand~5 & WO2  \\
Westerlund~1 &  13.5 & 10& 4.5$\pm$0.5&(40::) & WR77e,q,sd& J,R,X 
&WN5o,h\\
         &        &&&    & WR77a,s,sa,sb& O,Q,U,W &WN6o,h\\
         &        &&&    & WR77d,j,o,r,sc& A,B,D,G &WN7b,o     \\
         &        &&&    & WR77c,h&I,V & WN8o                 \\
         &        &&&    & WR77k& L & WN9                 \\
         &        &&&    & WR77f& S & WN10-11:                \\
         &        &&&    & WR77g& K &WC8                   \\
         &        &&&    & WR77b,i,l,m,n,p,aa& C,E,F,H,M,N,T & WC9\\
Pismis~20 & 12.7 & 8& 4.5: &  40 & WR67 & LSS~3329 & WN6        \\ %
NGC~6871        & 11.65& 7 & 3.2$\pm$0.7& 40 & WR133 & HD~190918 & 
WN5+O9.5Ia\\
Berkeley~86     & 11.4 & 7 & 3.7$\pm$1 & 35 & WR139 & V444 Cyg & 
WN5+O6 \\
Bochum~7    &  13.45 &  9 & 2.8$\pm$0.5 &  35 & 
WR12 & LSS~1145 &  WN8h \\
Ruprecht~44    & 13.35 &  8 &3.5 & 22  & WR10 & HD~65865 &  WN5h \\
Markarian 50 & 12.8 & 8&  7--10 & 18 & WR157 & HD~219460B  & WN5\\
\hline
\end{tabular}
\end{center}
 (1) Humphreys 1978; (2) Moffat 1983; (3) Perry et al. 1991; (4) Moffat et al. 
1991; (5) Massey \& Thompson 1991; (6)
 Massey \& Johnson 1993; (7) Massey et al. 1995; (8) Massey et al. 2001; 
 (9) Corti et al. 2003; (10) this study
\end{table*}

\section{WR properties}\label{properties}

\subsection{Binarity}

From the known WR sample  within Westerlund 1, none have been subject 
to photometric or 
spectroscopic monitoring, hindering efforts towards establishing a binary
fraction. OB companions for Milky Way WR stars have typically been 
suspected from the presence of photospheric absorption lines in blue 
optical spectroscopy, which is unrealistic for highly reddened Westerlund~1 
members. Nevertheless, two methods are at our disposal with regard to
investigating WR binarity. 

First, persistent and episodic dusty  WC stars are considered
to be in binary systems with OB companions (e.g. WR104, Tuthill et al. 
1999). Hydrogen from the OB star and carbon from the WC star provide 
 suitable ingredients, whilst their wind interaction regions 
may provide shielding from their intense  UV radiation fields plus 
the necessary high densities (e.g. Crowther 2007).

With regard to the presence or absence of dust within WC stars, 
T, N, C, H, M and F all reveal a strong near-IR excess with respect to the 
normal stellar energy distribution (recall Fig.~\ref{nir})
indicative of hot dust (Williams et al. 1987). K-band spectral
features are strongly diluted due to the dust with respect to shorter
wavelengths where the contribution of dust is reduced 
(Fig.~\ref{WCatlas}). Consequently at least 
75\% of the WC stars are likely to be binary systems 
from the presence of dust emission. 

For stars K (WC8) and E (WC9) the situation is less clear. The ratio of 
C\,{\sc iii} 2.110$\mu$m/0.971$\mu$m in star K is identical to the 
non-dusty WC8 star WR135, such that K does not appear to be forming dust. 
Similarly, the ratio of C\,{\sc iii} 2.110$\mu$m/0.971$\mu$m in
star E is similar to the non-dusty WC9 star WR75c  (Hopewell et al.
2005). In addition, the {\it Spitzer} 
IRAC 3.6--8$\mu$m colours  of star E presented in Table~\ref{IRAC}
are similar to non-dusty WN stars,  so it too is probably not forming
dust. Of course, the absence of dust emission does not exclude binarity,
since relatively short period late-type WC binaries (e.g. $\gamma$ Vel, 
WC8+O) are known not to form dust. Consequently, the WC binary fraction 
could be as high as 100\%.

Second, WR stars in binary systems with a
massive companion are often hard X-ray emitters (e.g. V444
Cyg, WN5+O), produced in the shocked region where their winds collide.
Most single OB and WN stars are relatively soft in X-rays (kT = 0.5 keV,
Skinner et al. 2002), whilst those in close binaries additionally possess
a harder X-ray component of $\sim$2--3 keV, presumably from the shocked
wind-wind collision zone.

Westerlund~1 has been observed with {\it Chandra}, from which Skinner et 
al.  (2006) and Clark et al. (2006) have matched strong, hard X-ray 
sources to  the known Wolf-Rayet population. WN stars A, B and L,
plus the dust forming WC9 star F 
possess strong ($L_{X} \geq 10^{32}$ erg\,s$^{-1}$), 
hard X-ray emission with a characteristic temperature of 2--3keV, plus
radio detections. In addition, several other WN stars possess hard
X-ray components, albeit with typically lower $L_{X}$ and no radio
detections, namely D, W, G, R, O and U.
Consequently, 3 out of 16 of the WN stars are firm binary candidates,
whilst another 6 are possible WN binaries from the hardness of their 
X-ray colours. Recall several He\,{\sc i} lines in star A exhibit central 
P Cygni 
absorption,  which is also suggestive of an early type companion. Note the 
weak  
emission line strength of star B led Negueruela \& Clark (2005) to suggest 
that this was a binary.

Therefore, a significant  fraction of the  WR stars in Westerlund~1 appear 
to be 
members of close, massive binary systems, with a minimum of
3/16 WN stars (X-rays) plus 6/8 WC stars (dust, X-rays), i.e. 
$\geq$38\% of the total WR content, or more likely a total of 9/16 WN
stars plus 6/8 WC stars i.e. $\geq$62\% of the total WR content,
including the other hard X-ray sources. However, to reiterate, 
our present observations do not exclude a binary fraction of 100\%.

\subsection{Current stellar masses}

There is a well known mass-luminosity relationship for H-deficient
WR stars (e.g. Schaerer \& Maeder 1992) from which we may estimate
current masses. For our derived distance modulus  of 13.5 mag, plus
individual K$_{S}$-band
extinctions, we may 
determine the absolute K$_{S}$-band magnitude $M_{\rm K_{S}}$,
which is representative of the stellar continuum, except for the dusty
WC stars. Quantitative analysis of individual WR stars in Westerlund~1 is 
presently in progress. Until such results are available, calibrations
for individual subtypes (e.g. Crowther 2007) need to be applied.
WR bolometric corrections are typically quoted for the $v$-band,
such that theoretical $(v - K_{S})_{0}$ WR spectral energy distribution colours 
also need to be used  (see Table~A1).
Results from this approach are presented in 
Table~\ref{masses} and suggest current WR masses in the range 
10--18$M_{\odot}$ with 14$M_{\odot}$ on average. These are fairly
typical of Milky Way WR masses, as deduced from binary orbits 
(e.g. Crowther  2007).

We have excluded the dusty WC9 stars from this sample,
since dust contributes to their K$_{S}$-band magnitudes, such that bolometric
magnitudes and masses would  otherwise be seriously overestimated. If 
we were to
apply dust-free WC9 intrinsic colours from Table~A1 for the 
six WC9d stars to derive $A_{\rm K_{S}}$, we would obtain  --7.0
$\leq$ $M_{\rm K_{S}}$ $\leq$ --9.1  mag. Application of standard 
bolometric
corrections would naturally yield unrealistic current masses in the 
range 30--180 $M_{\odot}$. 

Consequently, great care has to be taken when 
using near-IR observations of dusty WC stars as indicators of distance or
stellar luminosity. For example, Eikenberry et al. (2004) justify
a large distance towards the SGR~1806-20 cluster partially on the basis of 
$M_{K_{S}}$ =  --8.6 mag for their `star B' with a 
WC9d spectral type. 
Weak K-band spectral features are observed in this star, one would 
expect only moderate dust contamination, i.e. a K$_{S}$-band absolute 
magnitude close to --7 mag, and so a  near distance, broadly 
consistent with that suggested by Cameron et al. (2005).

\begin{table}
\caption{Mass estimates of WR stars for our deduced distance 
modulus of 13.5 mag, plus bolometric corrections from the 
calibration of Crowther (2007) and theoretical colours from
Table 8, together with the mass-luminosity relation for
H-free WR stars (Schaerer \& Maeder 1992).}
\label{masses}
\begin{tabular}{
c@{\hspace{2mm}}
l@{\hspace{2mm}}
c@{\hspace{2mm}}
c@{\hspace{2mm}}
c@{\hspace{2mm}}
c@{\hspace{2mm}}
c@{\hspace{2mm}}
l}
\hline
Star & Sp Type  & $M_{\rm K_{S}}$ & $M_{\rm bol} - M_{v}$ & $M_{v} - M_{\rm K_{S}}$ 
& $M_{\rm bol}$ & $M$ \\
     &          & mag         & mag                   & mag  
& mag           & $M_{\odot}$ \\
\hline 
A    & WN7b & --6.0 & --4.1 &   0.6 & --9.5 & 18.3 \\
B    & WN7o & --5.3 & --3.7 & --0.2 & --9.2 & 15.3 \\
D    & WN7o & --5.0 & --3.7 & --0.2 & --8.9 & 13.3 \\
G    & WN7o & --5.4 & --3.7 & --0.2 & --9.3 & 16.0 \\
I    & WN8o & --5.8 & --3.2 & --0.2 & --9.2 & 15.3 \\
J    & WN5h & --4.8 & --4.2 & --0.2 & --9.1 &    \\
L    & WN9h:& --7.2 & --2.8 & --0.2 & --10.3 &   \\
O    & WN6o & --4.8 & --4.2 & --0.2 & --9.2 & 14.9 \\
P    & WN7o & --5.2 & --3.7 & --0.2 & --9.1 & 14.5 \\
Q    & WN6o & --4.4 & --4.2 & --0.2 & --8.8 & 12.3 \\
R    & WN5o & --4.1 & --4.2 & --0.2 & --8.4 & 10.1 \\
S    & WN10-11h: 
            & --6.0 & (--2.2) & (0.0) & (--8.2) & \\
U &   WN6o & --5.0 & --4.2 & --0.2 & --9.4 & 17.2 \\
V &   WN8o  & --5.8 & --3.2 & --0.2 & --9.2 & 15.8 \\
W &  WN6h & --4.5 & --4.2 &--0.2 & --8.9 & \\
X & WN5o & --4.5 & --4.2 & --0.2 & --8.8 &12.4 \\
\\
E & WC9 & --6.1 & --3.0 & --0.1 & --9.2 & 15.4 \\
K & WC8  & --4.9 & --4.1 & 0.5 & --8.5 & 10.5   \\
\hline
\end{tabular}
\end{table}

\section{Single versus binary evolution}\label{evol}

\subsection{Single star models}

We have compared the number, and subtype distribution, of WR stars
in Westerlund~1 with those predicted by the evolutionary synthesis model
Starburst99 (Leitherer et al. 1999), assuming Solar 
metallicity, an instantaneous burst, a Kroupa IMF and an upper 
mass limit of $\sim 120M_{\odot}$. Two Geneva group evolutionary models 
may be used, with standard or enhanced mass-loss rates for individual stars 
(Schaller et al. 1992; Meynet et al 1994). The latter is normally 
preferred, since rotation is neglected in such models. The total WR 
population is reasonably  well reproduced,  particularly for  the enhanced 
mass-loss case, although the ratio  N(WC)/N(WN) is far too  high in all  
cases. Indeed, WN stars are predicted to be H-rich (WNL from
an evolutionary definition), in contrast to the 
observed H-poor WN stars  within Westerlund 1. The number of O stars 
in  Westerlund~1 
-- predicted to  be N(O)=150--320 for an age of 4--5Myr --  awaits the 
completion of an ongoing near-IR  spectroscopic survey.

Table~\ref{eldridge} presents predicted WR subtype distributions from single
star evolution at ages of 4--5Myr from Eldridge (2006, priv. comm.), 
based upon Solar metallicity models
presented by Eldridge \& Tout (2004), in which  rotation is neglected, 
whilst convective  overshooting is considered. The  Eldridge \& Tout
models are favoured with
respect to  evolutionary models from Meynet et al. (1994) 
since their predicted N(WC)/N(WN) ratio is less extreme,  in better 
agreement with the Westerlund 1 Wolf-Rayet population. However,
theoretical N(WN H-rich)/N(WN H-poor)  
ratios remain larger than those observed in Westerlund~1. Note that
for an age of $\sim$4.5Myr, the WR progenitor mass lies in the range
40--55$M_{\odot}$.
As a result, it is likely that the progenitor of the pulsar
identified by Muno et al. (2006) had an initial mass in excess of 
$50M_{\odot}$.

Rotation is, thus far, neglected in such synthesis models, but has 
recently been  considered for  individual evolutionary models by Meynet \& 
Maeder (2003), from which some conclusions may be 
inferred. The impact of rotational mixing is to extend the WR lifetime and
reduce the mass threshold for WR formation, from 37 to 22$M_{\odot}$ for 
nominal initial rotation rates at Solar metallicity. This has the effect 
of increasing the N(WR)/N(O) number ratio, and decreasing the 
predicted N(WC)/N(WN) ratio. This would improve the  comparison with 
Westerlund~1. However, WN stars in Westerlund~1 appear to be largely 
H-poor, in contrast  with the extended H-rich WN phases and 
diminished H-poor WN phases  predicted by rotating models 
(Meynet \& Maeder 2003). 

\subsection{Binary evolution}

Close binary evolution has been neglected thus far,
so we additionally need to consider this formation channel
for WR stars, in particular given the high density within the cluster 
core. We know of at least one pulsar in Westerlund~1 which has 
probably been formed  as a  result of massive close binary evolution (Muno 
et al.  2006).

\begin{table}
\caption{WR number ratios predicted by single and binary
evolutionary models at Solar metallicity 
from Eldridge (2006, priv. comm.), based upon models presented
by Eldridge \& Tout (2004)  in which rotational mixing
is neglected, versus the observed WR population in Westerlund~1. 
Similar results for single stars are obtained by Meynet et al. 
(1994). Eldridge \& Tout (2004) define an upper limit to a H-poor WN star 
as X(H)=0.1\%, by mass,  whilst the approximate observational limit is $\sim$5\%.
}\label{eldridge}
\begin{tabular}{lccc}
\hline
Age & N(WN H-rich)/  & N(WC)/ & Mass Range \\
Myr & N(WN H-poor)  &  N(WN)  & $M_{\odot}$ \\
\hline
\multicolumn{4}{c}{Single stars} \\
4--4.5 & 0.8 & 1.5 & 45--60 \\
4.25--4.75 & 1.2 & 2.4 & 40--55 \\
4.5--5.0 & 1.5 & 1.6 & 38--45 \\
\\
\multicolumn{4}{c}{Binary stars}\\
4.0--4.5 & 2.0 & 0.9 & \\
4.25--4.75 & 1.1 & 1.4 & \\
4.5--5.0 & 0.7 & 1.1 & \\
\\
\multicolumn{4}{c}{Westerlund~1} \\
        & 0.25 & 0.5 & \\
\hline
\end{tabular}
\end{table}

Theoretically, 
close binary evolution causes the premature loss of the H-rich mantle
during Roche lobe overflow or common envelope evolution, leading to  
an extended H-rich WN phase. Unfortunately, 
predictions greatly depend upon a large number of additional parameters, 
notably the initial binary fraction, their mass ratio and their initial 
period distribution. 
Nevertheless, the predicted WR subtype distribution for close
binary is presented in Table~\ref{eldridge}, which improves the predicted
N(WC)/N(WN) ratio, but does not resolve the discrepancy.

The location of two WR stars far from the central cluster of Westerlund~1
(T and N)  suggests either an ejection via dynamical interactions 
within the cluster at an early phase, or the recoil following a supernova
(SN) explosion within a massive binary system. Both lie 4.5 arcmin
from the cluster, which equates to a projected distance of $\sim$7~pc,
suggesting velocities of $\sim$1.5--6 km/s for ejection 1--4\,Myr ago. 
However, both systems are dust forming WC9 stars, suggesting
binary OB companions. Neither method of ejecting the WR from the cluster 
core outlined above would naturally explain such a scenario -- dynamical
interactions naturally favour ejection of single stars, whilst a runaway
nature would require that the companion has already undergone a SN 
explosion.

\section{Conclusions}\label{conc}

We present interference filter  Y- and K-band imaging and follow up near-IR
spectroscopy  in order to identify the complete WR
population within Westerlund~1. Our results are as follows:

\begin{enumerate}
\item Four WR stars in Westerlund~1 are newly confirmed here, with 
nomenclatures of U, V, W, X following Clark \& Negueruela (2002),
Negueruela \& Clark (2005) and Hopewell et al. (2005), bringing the
total number of WR stars to 24.
Of these, U and V were previously discovered from optical 
spectroscopy (Negueruela, 2005, priv. comm.),  whilst U, W and X 
have been independently reported by Groh et al. (2006). Indeed, our 
photometric  approach  is similar to Groh et al. (2006) except 
that  we probe  deeper in the Y-band and cover a larger field, but
neglect the He\,{\sc ii} 1.0124$\mu$m filter.

\item Intermediate resolution JHK spectroscopy  for all 24 WR stars in Westerlund~1 are 
presented. We present a quantitative near-IR classification scheme
for WN and WC stars, calibrated using line strengths measured in 
optically visible field WR stars. From our sample, 16 WN stars are fairly 
evenly divided between mid (WN5--6)  and late ($\geq$WN7) subtypes, 
whilst all 8 WC stars are late subtypes (7/8 are WC9 stars). The majority 
($\sim$75\%) of early and late WN stars are H-poor, as estimated from
the strength of Paschen/Brackett lines. Amongst the WN stars, solely
star A shows a strong, broad line spectrum (WN7b), which is indicative
of a dense, fast stellar wind.

\item
Dominant late WC
populations, such as those in Westerlund~1 are well known in metal-rich 
regions such as the  inner Milky Way (Hopewell et al. 2005) and M83 
(Hadfield et al. 2005). Early subtypes dominate in metal-poor 
regions. The reason for this dichotomy appears to be, at least in part,
due to  the sensitivity of 
WC spectral classification to (metallicity-dependent) wind strength 
(Crowther et al. 2002), rather than elemental abundance 
differences.

\item From a total of eight WC stars within Westerlund 1, six contain hot 
dust, which has the 
effect of strongly diluting the K-band spectral features with respect to 
shorter wavelengths - e.g. C\,{\sc iii} 2.11$\mu$m/0.971$\mu$m = 0.42 for 
the only apparent 
non-dusty WC9 star (E) versus 0.02 for star F. 2$\mu$m  narrow-band 
surveys  presently underway to identify visibly obscured 
WR stars in the Milky Way (e.g. Homeier et al. 2003ab)
will not be so sensitive to such dusty WC stars due to the dilution of
their emission lines by dust continuum.
Instead, dust forming WC stars will most easily be detected at 
shorter wavelengths 
where their stellar signature is much more prominent, or via their hot dust
signatures from near and mid-IR photometry. 

\item
It is widely believed that dust formation in WR stars requires
the presence of an OB companion, such that most WC stars within
Westerlund~1 are binaries. The presence of hard X-ray
emission from a subset of WN stars (Skinner et al. 2006; Clark et al. 2006) 
indicates a minimum WR binary fraction of $\sim$38\% in Westerlund~1,
or more likely 62\%. Indeed, 
our present observations do not exclude a WR binary fraction of 100\%.

\item We apply an absolute K-band calibration for our (non-dusty) WR
stars to estimate the distance and extinction, revealing $\sim$5.0
kpc and $A_{K_{S}}$=0.96 $\pm$0.14 mag, in reasonable agreement with 
previous studies based upon yellow hypergiants and pre-main sequence stars 
(Clark
et al. 2005; Brandner et al. 2006).  WC9 stars were excluded from our
calibration since non-dusty WC9 stars were not previously known within
optically visible clusters (van der Hucht \& Williams 2006). Applying our
Westerlund~1 distance to the only non-dusty WC9 star (star E), suggests
$M_{K_{S}}$ = --6.2 mag (the remainder span --7.0 to --9.1 mag).

\item For an assumed Solar galactocentric distance of 8.5 kpc,
Westerlund~1 lies at a galactocentric distance of 
4.2$^{+0.75}_{-0.35}$ kpc,
i.e.  similar to the outer edge of the Galactic bar (Benjamin et al.
2005). This location may prove significant for the formation of a 10$^{5}
M_{\odot}$ cluster within the Milky Way. Based on the Esteban et al.
(2005) Milky Way oxygen metallicity gradient, Westerlund~1 is anticipated
to be moderately oxygen-rich, exceeding that of the Orion Nebula by 60\%,
i.e.  $\log$ (O/H) + 12 $\sim$8.87.

\item For the hydrogen-deficient, non-dusty WR stars we have applied
model atmosphere bolometric corrections and intrinsic visual to JHK$_{S}$ 
colours to provide estimates of stellar luminosities, from which masses 
follow using the mass-luminosity relation of Schaerer \& Maeder 
(1992). We estimate current masses of 10--18$M_{\odot}$, typical of 
binary orbit deduced masses for Milky Way Wolf-Rayet stars (e.g. Crowther 2007).

\item  We provide revised mass limits for WR progenitors 
in other optically visible Milky Way clusters, updated from Massey et al. 
(2001). We adopt revised OB spectral type--temperature calibrations
and apply Solar metallicity isochrones from Meynet et al. (2004).
Westerlund~1 more than doubles the known statistics, and includes 
subtypes not previously covered. The observed ratio N(WR)/N(RSG + YHG) 
$\sim$3 for Westerlund~1 
suggests an age of $\sim$4.5--5.0~Myr, from comparison with evolutionary 
models,
such that the WR stars are descended from stars of  
initial mass $40-55 M_{\odot}$ 
with potential implication for a higher progenitor mass for 
the pulsar recently observed in Westerlund~1 (Muno et al. 2006). 
Consequently, approximately 75\% of the WR progenitor mass has
been removed due to stellar winds or close binary mass-transfer.

\item Comparisons between the observed WR population in Westerlund~1 and 
evolutionary models for rotating/non-rotating Solar metallicity stars
is rather poor, in the sense that the observed ratios of N(WC)/N(WN) and
N(WN H-rich)/N(WN H-poor) are rather lower than predicted, for both
single and binary models.
\end{enumerate}

We intend to follow up the present observational study with quantitative
analysis of the WR population in Westerlund~1 from which we will
be able to extract more reliable abundances, stellar luminosities 
and current masses, plus wind properties, with which to better test
current evolutionary predictions for single and binary massive stars.

\section*{acknowledgements} Thanks to Cedric Foellmi for obtaining
jittered J, H and K$_{S}$ band NTT/SOFI images of Westerlund~1, to Nicole
Homeier for providing NTT/SOFI spectroscopy of comparison broad-lined WN
stars, to John Eldridge for generating comparisons with evolutionary
predictions, and to Jose Groh for distributing his preprint prior to
publication. Malcolm Currie provided invaluable help with applying ORAC-DR
to ESO datasets.  We appreciate critical comments from the referee,
Peredur Williams, which helped improve several aspects of this paper. PAC
acknowledges financial support from the UK Royal Society. IN is a
researcher of the programme {\em Ram\'on y Cajal}, funded by the Spanish
Ministerio de Educaci\'on y Ciencia (MEC) and the University of Alicante,
with partial support from the Generalitat Valenciana and the European
Regional Development Fund (EDRF/FEDER). This research is partially
supported by the MEC under grant AYA2005-00095. This publication
makes use of data products from the Two Micron All Sky Survey, which is a
joint project of the University of Massachusetts and the Infrared
Processing and Analysis Center/California Institute of Technology, funded
by the National Aeronautics and Space Administration and the National
Science Foundation

\clearpage

\noindent {\bf APPENDIX A: Absolute magnitudes and intrinsic colours of 
Wolf-Rayet stars}\label{appendix}

In this appendix, we provide a calibration of absolute K$_{S}$-band magnitudes for (primarily)
cluster or association WN  and  WC stars 
based upon 2MASS photometry and synthetic J,H,K$_{S}$ colours. The 
latter were obtained
from convolving 
CMFGEN (Hillier \& Miller  1998)
stellar atmosphere models of individual WR stars with SOFI J, H, 
K$_{S}$ filter
response profiles, with zero points defined by Vega, as is usual.
As such, our approach complements the recent empirical JHKL' photometric 
study for WR stars of van der Hucht \& Williams (2006).
Dust forming WC stars are excluded 
from this analysis since their near-IR colours and absolute magnitudes possess a large scatter, with
(J-H)$_{0}$ = 1$\pm$0.7 mag, (H-K)$_{0}$ = 1.1$\pm$0.6 mag and $M_{\rm 
K}$=--8.5$\pm$ 1.5 mag according to van der Hucht \& Williams (2006).

In common with the method outlined 
within Section~\ref{distance}, interstellar
K$_{S}$-band extinctions are estimated independently using the J--K$_{S}$ and 
H--K$_{S}$ colours of individual cluster/association stars, adopting the 
Indebetouw et al. (2005) relationship  between J, H and K$_{S}$ 
extinction. We used the average of the two  approaches for our
final K$_{S}$-band extinction, $\overline{A_{K_{S}}}$. Absolute magnitudes
follow from previously determined cluster/association distance moduli
(van der Hucht \& Williams 2006).

We have grouped  early WN stars according to weak versus strong lined stars, since these
stars possess rather different near-IR colours, such that the strong lined stars
possess rather stronger stellar winds, and so flatter (free-free) near-IR energy distributions.
Our  adopted near-IR calibration is presented in Table~A1.

\begin{table*}
{\bf Table A1:} Calibration of 
absolute K$_{S}$-band magnitudes and intrinsic J,H,K$_{S}$ colours
for Galactic WN3--9 and non-dusty WC5--9 stars, supplemented by two LMC WN10-11 stars, 
based upon 2MASS photometry and theoretical stellar atmospheric model continua for 
individual stars. Cluster  or  association distances are taken from van der Hucht \& Williams
(2006).  A single $M_{K_{S}}$ is  given for WN7--9 stars due to the low number of cluster 
or association members (WN7ha stars are omitted from the present calibration) whilst $M_{K_{S}}$ is 
absent for WC9 stars since neither WR88 nor WR92 are known cluster or association members.
Adopted values for each group are shown in bold.
\begin{tabular}{
l@{\hspace{3mm}}
l@{\hspace{3mm}}
r@{\hspace{3mm}}
r@{\hspace{3mm}}
r@{\hspace{3mm}}
r@{\hspace{3mm}}
r@{\hspace{3mm}}
r@{\hspace{3mm}}
r@{\hspace{3mm}}
r@{\hspace{3mm}}
r@{\hspace{3mm}}
c@{\hspace{3mm}}
r}
\hline
Star & Subtype 
& (v-J)$_{0}$ & (J-K$_{S}$)$_{0}$ & (H-K$_{S}$)$_{0}$ & J--K$_{S}$ & H--K$_{S}$ & 
$A^{J-K_{S}}_{K_{S}}$ &  $A^{H-K_{S}}_{K_{S}}$ 
& $\overline{A_{K_{S}}}$ & $K_{S} - \overline{A_{K_{S}}}$ & DM & $M_{K_{S}}$ \\
DM & 
$M_{\rm K}$ \\
     &         & mag         & mag         & mag         & mag & mag & mag & mag & mag & mag & mag & mag \\
\hline
WR46      & WN3p & --0.78 &--0.11&--0.03& 0.37 & 0.25 & 0.32 & 0.51 & 0.41 & 9.42 & 13.05 &--3.63   \\
WR152     & WN3  & --0.78 &--0.11&--0.03& 0.45 & 0.28 & 0.37 & 0.56 & 0.47 & 9.57 & 12.2  & --2.63  \\
\multicolumn{2}{l}{\bf WN3--4 (weak)} 
                &        &{\bf --0.11} &{\bf --0.03}&      &      &      &      &      &      &       & {\bf --3.13}\\ 
\\
WR10      & WN5h & --0.56 & 0.04 & 0.07 & 0.44 & 0.28 & 0.27 & 0.39 & 0.33 
& 9.28 & 13.35 & --4.07\\
WR115     & WN6  & --0.31 & 0.17 & 0.15 & 1.05 & 0.48 & 0.59 & 0.61 & 0.60 & 6.34 & 11.5  & --5.16\\
WR138     & WN5+?& --0.23 & 0.21 & 0.17 & 0.38 & 0.22 & 0.11 & 0.08 & 0.10 & 6.38 & 10.5  & --4.02\\
WR141     & WN6+O?& --0.31 & 0.17 & 0.15 & 0.81 & 0.40 & 0.43 & 0.46 & 0.44 & 6.09 & 10.5  & --4.41\\ 
\multicolumn{2}{l}{\bf WN5--6 (weak)} 
                &        &{\bf  0.18} &{\bf  0.16} &      &      &      &      &      &      &       & {\bf --4.41}\\ 
\\
WR1       & WN4b& 0.29   & 0.38 & 0.27 & 0.73 & 0.38 & 0.23 & 0.20 & 0.22 & 7.26 & 11.3  & --4.04\\
WR134     & WN6b& 0.23   & 0.37 & 0.27 & 0.55 & 0.36 & 0.12 & 0.17 & 0.14 & 6.02 & 11.2  & --5.18 \\
WR136     & WN6b& 0.21   & 0.36 & 0.26 & 0.57 & 0.34 & 0.14 & 0.14 & 0.14 & 5.42 & 10.5  & --5.08 \\
\multicolumn{2}{l}{\bf WN4--7 (strong)}  
               &         & {\bf 0.37} &{\bf  0.27} &      &      &      &      &      &      &       & {\bf --4.77}\\ 
\\
WR78      & WN7h &--0.43 & 0.09 & 0.09 & 0.46 & 0.29 & 0.24 & 0.37 & 0.31 & 4.67 & 11.5  & --6.83\\
WR40      & WN8h &  0.06 & 0.28 & 0.19 & 0.51 & 0.30 & 0.15 & 0.21 & 0.18 & 5.93 \\
WR66      & WN8+?&--0.47 & 0.07 & 0.07 & 0.78 & 0.33 & 0.48 & 0.47 & 0.47 & 7.68 & 12.57 & --4.89\\
WR105     & WN9h &--0.56 & 0.07 & 0.10 & 1.30 & 0.51 & 0.82 & 0.75 & 0.78 & 4.95 & 11.0  & --6.05 \\
\multicolumn{2}{l}{\bf WN7--9 (Weak)}  
                 &       & {\bf 0.13} & {\bf 0.11} &      &      &      &      &      &      &       &{\bf --5.92} \\ 
\\
BE294     & WN10h&--0.09 & 0.29 & 0.20 & 0.40 & 0.29 & 0.07 & 0.17 & 0.12 & 11.52 & 18.45 & --6.93 \\
S119      & WN11h&--0.35 & 0.07 & 0.05 & 0.10 & 0.13 & 0.02 & 0.15 & 0.08 & 11.64 & 18.45 & --6.81 \\
\multicolumn{2}{l}{\bf WN10--11}
                &        &{\bf 0.18} & {\bf 0.12} &       &     &      &      &      &       &       & {\bf --6.87} \\ 
\\
WR111     & WC5 &0.09    & 0.66 & 0.59 & 0.77 & 0.63 & 0.07 & 0.07 & 0.07 & 6.44 & 11.0  & --4.56 \\           
WR154     & WC6 &0.06    & 0.61 & 0.59 & 1.01 & 0.72 & 0.26 & 0.24 & 0.25 & 8.04 & 12.2  & --4.16 \\
WR14      & WC7 & 0.09   & 0.59 & 0.57 & 0.88 & 0.63 & 0.19 & 0.10 & 0.15 & 6.46 & 11.5  & --5.04 \\
\multicolumn{2}{l}{\bf WC5--7}   
                &        & {\bf 0.62} &{\bf 0.58} &      &      &      &      &      &      &       & {\bf --4.59} \\ 
\\
WR135     & WC8 & 0.07   & 0.43 & 0.38 & 0.57 & 0.45 & 0.09 & 0.12 & 0.11 & 6.55 & 11.2  & --4.65 \\
\multicolumn{2}{l}{\bf WC8} 
                &        &{\bf 0.43} & {\bf 0.38} &      &      &      &      &      &      &       &{\bf --4.65}\\ 
\\
WR88      & WC9 & --0.29 & 0.23 & 0.26 & 0.98 & 0.51 & 0.50 & 0.46 & 0.48 & 7.57 \\
WR92      & WC9 & --0.29 & 0.23 & 0.26 & 0.68 & 0.40 & 0.30 & 0.26 & 0.28 & 8.54 \\
\multicolumn{2}{l}{\bf WC9} 
                &        &{\bf 0.23} & {\bf 0.26} &      &      &      &      &      &      \\         
\hline
\end{tabular}
\end{table*}

\clearpage

\noindent {\bf APPENDIX B: Galactic cluster turn-off masses}

In Table~B1 we list properties for the highest (initial) mass
OB stars within Milky Way clusters which contain WR stars. We 
follow the approach of from Schild \& Maeder (1984) and Massey et al. 
(2001), updated to account for the revised spectral  type-$T_{\rm 
eff}$ calibration of  Martins et al. (2005) for O stars, Crowther et al. 
(2006)  for B supergiants,  plus the high mass-loss Meynet et 
al. (1994) Solar metallicity isochrones. A summary of cluster turn-off 
masses is shown in Table~\ref{cluster}. 

\clearpage

\begin{table*} 
{\bf Table B1:} Parameters derived for the highest mass
unevolved stars within selected optically visible Milky Way clusters.
Parameters for a distance of 2.6~kpc towards the Carina~OB1 clusters are
listed here, following Humphreys (1978), is preference  to
the distance of 3.2~kpc derived by Massey \& Johnson (1993).
\begin{center}
\begin{tabular} { l@{\hspace{3mm}} l@{\hspace{3mm}} r@{\hspace{3mm}}
c@{\hspace{3mm}} c@{\hspace{3mm}} c@{\hspace{3mm}}
l@{\hspace{3mm}} l@{\hspace{3mm}} c@{\hspace{3mm}} r@{\hspace{3mm}}
r@{\hspace{3mm}} c } 
\hline 
Star & Sp Type & V & B--V & E(B--V) & $M_{V}$ & Ref & $T_{\rm eff}$
& $\log L$ &Mass & Age\\
     &         & (mag)   &(mag) & (mag)&(mag)& & (kK)  & 
$(L_{\odot}$)&($M_{\odot}$) & (Myr)\\
\hline
\multicolumn{8}{c}{\bf Cyg~OB2 (DM=11.2 mag)}\\
Cyg~OB2 22 & O4\,III & 11.55 & 2.04 & 2.36 & --7.0 & 5 & 42.0 & 6.22 &114 
&1.8\\
Cyg~OB2 8a & O6\,Ib  & 9.06  & 1.30 & 1.60 & --7.1 & 5 & 37.0 & 6.13 &97 
&2.0\\
Cyg~OB2 9  & O5\,If  & 10.96 & 1.81 & 2.11 & --6.8 & 5 & 38.5 & 6.05 
&83&2.0\\
Cyg~OB2 11 & O5\,If+ & 10.03 & 1.49& 1.79 & --6.7 & 5 & 38.5 & 6.02 & 80 
&2.0\\
\\
\multicolumn{8}{c}{\bf NGC~3603 (DM=14.25 mag)}\\
HD~97950-A2& O3\,V &12.0: & 1.09: & 1.41 & --6.6 &7& 45 & 6.16 & 103 & 
1.3\\
HD~97950-A3& O3\,III&12.5: & 1.09: & 1.41 & --6.1 &7& 44.5&5.95 & 75 & 
1.1\\
MDS~42     & O3\,III&12.8: & 1.09:& 1.41 & --5.8 &3& 44.5&5.83 & 64 & 
1.1\\
MDS~24     & O4\,V  &12.69:& 1.09: & 1.41 & --5.9 &3& 43  &5.83 & 62 & 
1.6\\
\\
\multicolumn{8}{c}{\bf Pismis 24 (DM=12.0 mag)}\\
HDE~319718 & O3.5\,If & 10.43 & 1.45 & 1.75 & --7.0 & 9,10& 41 & 6.20 & 
108&1.8\\
Pis~24--17 & O3.5\,III& 11.84 & 1.49 & 1.81 & --5.8 & 9,10& 43 & 5.76 & 57 
& 1.5\\
Pis~24--2  & O5.5\,V  & 11.95 & 1.41 & 1.73 & --5.4 & 9  & 40 & 5.55 & 44 
& 2.0\\
Pis~24--13 & O6.5\,V  & 12.73 & 1.48 & 1.80 & --4.8 & 9  & 38 & 5.26 & 33 
& 2.0\\
\\
\multicolumn{8}{c}{\bf Trumpler~16 (DM=12.1 mag)} \\
HD~93250 & O3.5\,V&7.41 & 0.17 & 0.49 & --6.2 &6,10& 45  & 6.00& 80 & 1.1\\
HD~93205 & O3.5\,V&7.76 & 0.08 & 0.40 & --5.6 &6,10& 45   &5.75& 59 & 1.0\\
HDE~303308&O4\,V  &8.19 & 0.14 & 0.46 & --5.3 &6  & 43   &5.60& 48 & 1.3\\
HD~93204  & O5\,V &8.48 & 0.09 & 0.41 & --4.9 &6  & 37   &5.53& 37 & 1.3\\
\\
\multicolumn{8}{c}{\bf Havlen--Moffat 1 (DM=12.0 mag)}\\
LSS 4067         & O4\,If& 11.16 & 1.54 & 1.84 & --6.5 & 9  & 40 & 6.00 & 
77 & 1.8\\
C1715--387 No.~6 & O5\,If& 11.64 & 1.54 & 1.84 & --6.1 & 9  &38.5&5.76 & 
55 & 2.3\\
C1715--387 No.~8 & O5\,V & 12.52 & 1.52 & 1.84 & --5.2 & 9& 41  & 5.48&  
40 & 1.8\\
C1715--387 No.~12& O6\,If& 12.57 & 1.52 & 1.82 & --5.1 & 9& 39  & 5.32&  
33 & 2.7\\
\\
\multicolumn{8}{c}{\bf NGC~6231 (DM=11.5 mag)}\\
HD~152234 & B0\,Iab&5.44&0.20 & 0.44 & --7.4 &1,2 &27.5& 5.90 & 75: & 3.2:
\\
HD~152233 & O6\,III&6.56 & 0.14 & 0.46 &  --6.4 &1,2 &38  & 5.87 & 63  & 
2.2\\
HD~152249 & O9\,Iabp&6.44&0.20 & 0.48& --6.5 &1,2 &31  & 5.70 & 48 & 3.2\\
\\
\multicolumn{8}{c}{\bf Collinder~228 (DM=12.1 mag)}\\
HD~93206 & O9.5\,I & 6.28 & 0.14 & 0.41 & --7.1 & 9 & 30.5 & 5.90 & 63
&2.7\\
HD~93632 & O5\,III &8.39 & 0.29 & 0.61  & --5.6 & 9& 40.5 & 5.64 & 48
&2.0\\
HD~93130 & O7\,II  & 8.04 & 0.27 & 0.59 & --5.9 & 9& 36   & 5.61 & 44 
&2.7\\
HDE~305525& O6\,V  & 10.00 & 0.68 & 1.00 & --5.2 & 9& 39  &5.43&38 & 2.2\\
\hline
\end{tabular}
\end{center}
(1) Mermilliod 1976; (2) Levato \& Malaroda 1980;
(3) Moffat 1983;  (4) Moffat et al. 1991; 
(5) Massey \& Thompson 1991;  (6) Massey \& Johnson 1993;
(7) Moffat et al. 1994; (8) Massey et al. 1995; (9) Massey et al. 2001;
(10) Walborn et al. 2002; (11) Corti et al. 2003
\end{table*}

\begin{table*} 
{\bf Table B1: (continued)}
\begin{center}
\begin{tabular} { l@{\hspace{3mm}} l@{\hspace{3mm}} r@{\hspace{3mm}}
c@{\hspace{3mm}} c@{\hspace{3mm}} c@{\hspace{3mm}}
l@{\hspace{3mm}} l@{\hspace{3mm}} c@{\hspace{3mm}} r@{\hspace{3mm}}
r@{\hspace{3mm}} c } 
\hline 
Star & Sp Type & V & B--V & E(B--V) & $M_{V}$ & Ref & $T_{\rm eff}$
& $\log L$ &Mass & Age\\
     &         & (mag)   &(mag) & (mag)&(mag)& & (kK)  & 
$(L_{\odot}$)&($M_{\odot}$) & (Myr)\\
\hline
\multicolumn{8}{c}{\bf Trumpler~27 (DM=12.0 mag)}\\
Tr~27--27  & O8\,III & 13.31 & 2.16 & 2.47 & --6.3 & 9 & 34 & 5.73 & 51 & 
2.7 \\
LSS 4266  & B0.7\,Ia& 10.11 & 1.43 & 1.65 & --7.0 & 9 & 23 & 5.59 &  40 & 
4.2 \\
LSS 4253  & B0\,Ia  &10.55 & 1.28 & 1.52 & --6.2  & 9 & 27.5& 5.41& 35 & 
4.5 \\
\\
\multicolumn{8}{c}{\bf Westerlund~2 (DM=14.5 mag)}\\
Westerlund~2--18 & O7\,I &12.81 & 1.20 & 1.49 & --6.3 & 4 & 35 & 5.75 & 52
& 2.7 \\
Westerlund~2--157& O6\,V &14.14 & 1.36 & 1.68 & --5.6 & 4 & 39 & 5.58 & 44 
& 2.5\\
Westerlund~2--151& O6\,V &14.33 & 1.33 & 1.65 & --5.3 & 4 & 39 & 5.47 & 39 
& 2.5 \\
Westerlund~2--167& O7\,V & 14.19 & 1.29 & 1.61 & --5.3 & 4 & 37 & 5.41 & 
36 & 2.7 \\
\\
\multicolumn{8}{c}{\bf Berkeley~87 (DM=11.0 mag)}\\
HDE~229059 & B1\,Ia &8.71 & 1.52 & 1.71 &  --7.6 &9 & 21.5 & 5.76 & 49 & 
3.7: \\
Berk~87--25 & O8.5\,III & 10.46 & 1.29 &  1.60 & --5.5 & 9 & 33 & 5.35 & 
32 & 4.0 \\
Berk~87--4 & B0.2\,III & 10.92 & 1.26 & 1.53 & --4.8 & 9 & 28.5 & 4.98 & 21 
& 2.0\\
\\
\multicolumn{8}{c}{\bf Pismis~20 (DM=12.7 mag)}\\
HD~134959 & B2.5\,Ia & 8.20 & 0.93 & 1.08 & --7.8 & 9 & 16.5 & 5.63 & 39 & 
4.5 \\
Pis~20--2 & O8.5\,I  & 10.45 & 0.71 & 1.00 & --5.4 & 9 & 32 & 5.26 & 29 & 
4.5 \\
\\
\multicolumn{8}{c}{\bf NGC~6871 (DM=11.65 mag)}\\
HD~226868 & O9.7\,I & 8.81 & 0.83 & 1.09 &--6.2& 8& 30 & 5.60   & 38 &3.9 
\\
HD~190864 & O7\,III & 7.77:& 0.18: & 0.50 & --5.4& 8 & 36 & 5.43 & 36 &3.2
\\
HD~227018 & O7\,V   & 8.96 & 0.37 & 0.69 & --4.8 & 8 & 37 & 5.22 & 31 &2.5\\
\\
\multicolumn{8}{c}{\bf Berkeley~86 (DM=11.4 mag)}\\
HDE~228841 & O7\,V  & 8.98 & 0.60 & 0.92 & --5.3& 8 & 37 & 5.40 & 37 
&2.7\\
HD~228943  & B0\,V: & 9.30 & 0.87 & 1.17 & --5.7& 8 & 29.5& 5.31 & 30 & 
4.7:\\
HD~193595  & O8\,V  & 8.70 & 0.37 & 0.68 & --4.8& 8 & 28 & 5.15 & 28 
&3.5\\
HDE~228969 & O9.5\,V& 9.49 & 0.68 & 0.98 & --4.9& 8 & 32 & 5.10 & 25 & 
5.0:\\
\\
\multicolumn{8}{c}{\bf Bochum 7 (DM=13.45 mag)}\\
LSS 1131   & O7.5\,V & 10.80 & 0.51 & 0.82 & --5.2 & 11 & 36.0 & 5.33 & 33 
& 3.2 \\
LSS 1135 & O6.5\,V & 10.88 & 0.40 & 0.72 & --4.8 & 11 & 38.0 & 5.24 & 32 & 2.2 \\
LSS 1144 & O7.5\,V & 11.27 & 0.64 & 0.95 & --5.1 & 11 & 36.0 & 5.31 & 32 & 
3.2\\
\\
\multicolumn{8}{c}{\bf Ruprecht~44 (DM=13.35 mag)}\\
LSS~891 & O8\,III    & 10.93 & 0.29 & 0.60 & --4.3 & 9 & 34 & 4.90 & 22& 
3.5 \\
LSS~920 & O9.5\,V      & 11.38 & 0.24 & 0.54 & --3.6 & 9 & 31.5 & 4.56 & 
17 
& 3.5\\
\\
\multicolumn{8}{c}{\bf Markarian~50 (DM=12.8 mag)}\\
HD~219460A & B0\,III & 10.7: & 0.52 & 0.79 & --4.5 & 9 & 28.5 & 4.80 & 18
& 7.0: \\
Ma~50--31  & B0.5\,II& 11.21 & 0.64 & 0.86 & --4.3 & 9 & 26.0 & 4.60 & 15 
& 10.: \\
\hline            
\end{tabular}
\end{center}
(1) Mermilliod 1976; (2) Levato \& Malaroda 1980;
(3) Moffat 1983;  (4) Moffat et al. 1991; 
(5) Massey \& Thompson 1991;  (6) Massey \& Johnson 1993;
(7) Moffat et al. 1994; (8) Massey et al. 1995; (9) Massey et al. 2001;
(10) Walborn et al. 2002; (11) Corti et al. 2003
\end{table*}

\label{lastpage}
\end{document}